\newif\ifshaphered

\ifdefined\isshaphered
\shapheredtrue
\fi

\documentclass[10pt,journal,compsoc]{IEEEtran}

\ifCLASSOPTIONcompsoc
  \usepackage[nocompress]{cite}
\else
  \usepackage{cite}
\fi

\ifshaphered
\usepackage{longtable}
\newcommand{\revisionmajor}[1]{\textcolor{purple}{#1}}
\newcommand{\revision}[1]{#1}

\else
\newcommand{\revisionmajor}[1]{{#1}}
\newcommand{\revision}[1]{{#1}}

\fi

\newcommand*{\textlabel}[2]{%
  \edef\@currentlabel{#1}
  \phantomsection
  #1\label{#2}
}\usepackage{hyperref}

\usepackage[english]{babel}
\usepackage{blindtext}

\usepackage[utf8]{inputenc}

\usepackage{graphicx}

\usepackage{tikz}
\usetikzlibrary{spy}
\usetikzlibrary{arrows}
\usepackage{amsmath}
\usepackage{amsthm}
\usepackage{pifont}

\usepackage{algorithm}
\usepackage{algorithmicx}
\usepackage{multirow}
\usepackage{booktabs}
\usepackage{algpseudocode}
\usepackage{mdframed}

\usepackage{amsmath}
\usepackage{subfigure}
\usepackage{array}
\usepackage{enumitem}
\setlist[enumerate]{noitemsep, topsep=2pt}
\setlist[itemize]{noitemsep, topsep=2pt}
\setlist[description]{noitemsep, topsep=2pt, font=\normalfont\space}

\usepackage{pgfplots}
\pgfplotsset{compat=1.15} 
\pgfplotsset{%
    axis line origin/.style args={#1,#2}{
        x filter/.append code={ 
            \ifx\pgfmathresult\empty\else\pgfmathparse{\pgfmathresult-#1}\fi
        },
        y filter/.append code={
            \ifx\pgfmathresult\empty\else\pgfmathparse{\pgfmathresult-#2}\fi
        },
        xticklabel=\pgfmathparse{\tick+#1}\pgfmathprintnumber{\pgfmathresult},
        yticklabel=\pgfmathparse{\tick+#2}\pgfmathprintnumber{\pgfmathresult}
    }
}
\usepackage{xspace}
\usepackage{numprint}

\npthousandsep{,}
\newif\ifshowcomment
\showcommentfalse

\ifshowcomment
    \newcommand{\todo}[1]{\textsf{\color{red}{[{TODO: #1}]}}}
    \newcommand{\Jian}[1]{\textsf{\color{blue}{[{Jian: #1}]}}}
    \newcommand{\asokan}[1]{\textsf{\color{brown}{[{Asokan: #1}]}}}
    \newcommand{\ray}[1]{\textsf{\color{orange}{[{Ray: #1}]}}}
    \newcommand{\dawn}[1]{\textsf{\color{purple}{[{Dawn: #1}]}}}
    \newcommand{\bsy}[1]{\textsf{\color{green}{[{Bennet: #1}]}}}
    \newcommand{\pgao}[1]{\textsf{\color{cyan}{[{Peng: #1}]}}}
    \newcommand{\highlight}[1]{{\color{blue}{{#1}}}}
\else
    \newcommand{\todo}[1]{}
    \newcommand{\Jian}[1]{}
    \newcommand{\asokan}[1]{}
    \newcommand{\ray}[1]{}
    \newcommand{\dawn}[1]{}
    \newcommand{\bsy}[1]{}
    \newcommand{\pgao}[1]{}
    \newcommand{\highlight}[1]{#1}
\fi

\newif\ifATC
\ATCtrue

\ifATC
   \newcommand{\atc}[1]{} 
\else
   \newcommand{\atc}[1]{{#1}} 
\fi

\newcommand{\members}{\mathsf{EN}\textrm{s}\xspace}
\newcommand{\shards}{\textit{groups}\xspace}
\newcommand{\member}{\mathsf{EN}\xspace}

\newcommand{\Cli}{\mathsf{C}\xspace}
\newcommand{\Ser}{\mathsf{S}\xspace}
\newcommand{\order}{\mathsf{CN}\xspace}
\newcommand{\exec}{\mathsf{EN}\xspace}
\newcommand{\trans}{\textrm{TX}\xspace}
\newcommand{\st}{\textrm{ST}\xspace}

\newcommand{\res}{\textit{res}\xspace}

\newcommand{\sign}{\textsf{Sign}\xspace}
\newcommand{\verify}{\textsf{Verify}\xspace}
\newcommand{\aggregate}{\textsf{Aggre}\xspace}

\newcommand{\fig}{\textrm{Fig.}\xspace}

\newcommand{\remove}[1]{}

\newcommand{\cons}{\textrm{consensus}\xspace}
\newcommand{\name}{\textsf{SaberLedger}}
\newcommand{\BFT}{\textsf{Dagger}} 
\newcommand{\paradigm}{\textsf{Saber}}

\newcommand{\Shard}{execution group}

\makeatletter

\IEEEoverridecommandlockouts

\begin{document}

\title{Parallel and Asynchronous Smart Contract Execution}

\author{
Jian~Liu,~ 
Peilun~Li,~
Raymond~Cheng,~ 
N.~Asokan,~\IEEEmembership{Fellow,~IEEE,}
Dawn~Song,~\IEEEmembership{Fellow,~IEEE,}

\IEEEcompsocitemizethanks{
\IEEEcompsocthanksitem Jian Liu is with Zhejiang University, 

Raymond Cheng is with University of San Francisco, 

Peilun Li is with Tsinghua University,

N. Asokan is with University of Waterloo,

Dawn Song is with University of California, Berkeley.

E-mail: liujian2411@zju.edu.cn

lpl15@mails.tsinghua.edu.cn

me@raymondcheng.net

asokan@acm.org

dawnsong@cs.berkeley.edu 

}
}

\markboth{Journal of \LaTeX\ Class Files,~Vol.~14, No.~8, March~2020}%
{Shell \MakeLowercase{\textit{et al.}}: Bare Demo of IEEEtran.cls for Computer Society Journals}

\IEEEtitleabstractindextext{
\begin{abstract}
Today's blockchains suffer from low throughput and high latency,
which impedes their widespread adoption of more complex applications like smart contracts. 
In this paper, we propose a novel
paradigm for 
smart contract execution. 
It distinguishes between consensus nodes and execution nodes:
different groups of execution nodes can execute transactions in parallel;
meanwhile, consensus nodes can asynchronously order transactions and process execution results. 
Moreover, it requires no coordination among execution nodes and can effectively prevent livelocks. 
We show two ways of applying this paradigm to blockchains. 
First, we show how we can make Ethereum support 
parallel and asynchronous contract execution \emph{without hard-forks}. 
Then, we propose a new public, permissionless blockchain.
Our benchmark shows that, with a fast consensus layer, it can provide a high throughput 
even for complex transactions like Cryptokitties gene mixing. 
It can also protect simple transactions from being starved by complex transactions.

\end{abstract}

}

\maketitle

\ifshaphered

\onecolumn
\bgroup
\def\arraystretch{1.5}

\section*{Changelog}

We thank the reviewers and the editor for their detailed and insightful feedback. We have endeavored to address all their comments. 

In the following we list all received review comments in a table. In the 3rd column (``Response'') we explain how we addressed each comment. The 4th column (``Link'')  is a clickable link  that points to the place in the paper corresponding to the comment/response.
If a review is addressed at multiple locations in the paper, multiple links are provided. 
Each link points to to the text that is added (or changed) to address the corresponding comment.
Each link is named by the initial fragment of the text it points to.
To ease the tracking of changes, we marked changed text in \revisionmajor{purple}.

\subsection{Major Changes}



\begin{longtable}{p{.065\textwidth} | p{.35\textwidth} | p{.35\textwidth} | p{.1\textwidth}}
Reviewer & Review comment & Response & Link \\\hline

\textlabel{R2(1)}{row:r2_1}

&

OK, I agree with what you are saying, that it is fact that the ENs executing a transaction are a *random* selection across all ENs that makes it unlikely for an attacker to control the majority of them.

However, assuming an adversarial power of as low as 25\%, is an arbitrary and rather weak baseline. Blockchains can typically tolerate an adversarial power of just below 50\% (think of the "51\% attack"), which makes a tremendous difference to your attack assumptions. You should compute the probability of Equation 3 in Section 3.2 again, for a=50\%, and comment on it. If your model cannot support that high of an adversarial fraction, you should give an indication of what the EN size should be for 30\%, 40\%, 50\% adversarial power and give an explanation why this is not important. Aggregating 25\% of the mining power is not that difficult, if you are a popular mining pool.

&

Please be noticed that an adversarial power of 25\% is a common assumption for PoW blockchains, due to selfish mining.
If an adversary has more than 25\% power, the underlying identity chain is insecure. 

~

We have highlighted this in the paper.

~

We further remark that the probability of Equation 3 with $a=25\%$ is not proposed by ourselves, but by a sharding paper~\cite{Omniledger} published in a top-tier security conference.

&

\hyperref[r01:1]{Following prior work...}

\\\hline

\textlabel{R2(2)}{row:r2_2}

&

I do not agree: You added this sentence at the end of 5.1 vaguely saying that you could give some transaction fees to ENs, however this is not as simple as it sounds. It implies that ENs have to *trust* CNs on paying them the respective transaction fees out of good will. However, there's no obvious incentive for a CN to pass transaction fees on to ENs, and there's no obvious way for ENs to enforce that. In mainstream blockchains, the *only one* making a profit is a block's miner, and there's a reason for that: he is in control of who makes the profit, so he can be trusted to take it all.

&

Please be noticed that CNs run Byzantine consensus for every decision (including assigning transaction fees to ENs), so they can be {\em collectively} considered as a trusted party.

~

The clients who issue the transactions will pay the transaction fees, and CNs run Byzantine consensus to distribute the transaction fees according to the predefined rules.

&

\\\hline

\textlabel{R2(3)}{row:r2_3}

&

Thanks for fixing these. However, the paper still has several typos (e.g., ``cyrptokitties'') and lots of grammar errors. Even newly added text has some rather sloppy errors, like ``enables a some tasks'' or ``validity,t hey''! Careful editting by a native speaker would be absolutely necessary, should you proceed to resubmission.

&

Thanks for pointing out this. We have fixed these errors and have the paper proof-read by a native speaker.

&

\\\hline\hline

\textlabel{R3(1)}{row:r3_1}

&

Regarding group failures: My question is that even with a low probability, group failure very likely will happen sooner or later. Under the current design that separates consensus and smart contract execution, there is no mechanism to recover or mitigate group failures when they occur. Even in case of expected probability of $1/10^6$, it is certain that group failures will happen given the nature that blocks are continuously added to the chain. Considering millions or tens of millions of blocks can be generated, the likelihood of group failures is not negligible.

&

The group failure probability is {\bf independent} of the number of blocks being added to the chain.
After a group being constructed, with a probability of $1/10^6$, it is a ``faulty group'' (i.e., the number of failures is larger than $f$).
This probability will stay the same even with more blocks being added to the chain. 

~

An adaptive adversary may try to compromise more participants for a target group.
This is irrelevant to the group failure probability.
Following Hyperledger and other Sharding schemes, we periodically rotate the participants in each group to against such adaptive adversaries. 

~

This mechanism and probability is commonly used in sharding schemes~\cite{Omniledger, ZILLIQA}.

~

Thanks for pointing this out. We have added this discussion to the paper.
&

\hyperref[r02:1]{The group failure probability is independent...}

\\\hline

\textlabel{R3(2)}{row:r3_2}

&

Regarding extra latency. The question is about comparison with other sharding based smart contract execution approaches instead of comparison with non-sharding based smart contract systems. The locking design seems having a negative impact on latency, particularly for simple transactions assume that sharding is applied by default.

&

We did compare with other sharding based smart contract execution approaches.
The results can be found in Figure 6(a) and 6(b).
The performance of sharding are similar to $\name$ with \numprint{2000} TX/s,
but it requires 600 nodes in each shard, thereby \numprint{3467} nodes can only support 5 shards.

&

\\\hline

\textlabel{R3(3)}{row:r3_3}

&

Regarding monetary counter-incentive against live lock. As pointed out by other reviewer as well, I still have lingering doubt about this. I feel that this is perhaps one of the main ideas of this paper. However, it needs further clarification.

~

First, it is common that monetary counter-incentive is used as an approach to handle denial of service. Systems such as Hyperledger does not have gas fees but use similar design to separate smart contract execution from consensus. Could someone say, Hyperledger + monetary counter-incentive against live lock is equivalent in principle with SaberLedger?

~

Second, in this design, a malicious actor with resources to waste can still cause damage to the system even the attacker will run out of resources eventually. It is not clear how the system will function under the period of such attack.

&

First, adding monetary counter-incentive to Hyperledger will {\bf not} solve the adversarial livelock issue. 
See our example in the first tagged paragraph (i.e., ``Adversarial livelocks'') in Section 3.2:

``For example, two clients Alice and Bob share data objects $o_1$ and $o_2$ which are in two different groups $G_1$ (near Alice) and $G_2$ (near Bob) respectively.
Suppose Bob wants to make a transaction $\trans_B$ to update both $o_1$ and $o_2$ 
($\trans_B$ will first lock $o_2$ and then lock $o_1$). 
If Alice wants to make $\trans_B$ fail, she just needs to make a transaction $\trans_A$ to first lock $o_1$ and then lock $o_2$. 
In this case, both $\trans_A$ and $\trans_B$ will fail and Alice wins.
We name this attack {\em adversarial livelocks}.''

This can happen in Hyperledger as well, since Hyperledger adopts the ``execution-then-ordering'' paradigm. 
Same as the above example, Alice can make two execution groups in Hyperledger execute $\trans_A$ and $\trans_B$ in different order so that both transactions will fail and, {\bf Alice does not need to pay anything in this case (monetary conter-incentive will not apply).}
That means Hyplerleger cannot prevent such attack even with monetary counter-incentive.
In our scheme,  $\trans_A$ and $\trans_B$ will always be executed in the same order and monetary conter-incentive can be applied. 

~

Second, please be noticed that a malicious actor with resources to waste can do denial-of-service attack to any smart contract platform.  
For example, in Ethereum, a malicious actor with a fast network connection and willing to pay the gas fee, can always successfully call a smart contract and make other competing transactions fail.
However, the consistency property will always be maintained.
In this aspect, our scheme is equal to Ethereum.

~

We have added the above discussion to the paper. 

&

\hyperref[r02:3]{With monetary penalty...}

\\\hline

\textlabel{R3(4)}{row:r3_4}

&

Regarding comparison with related work. Judging questions from other reviewers, a constructive suggestion is to add a more clear comparison with related work, for instance using a taxonomy of designs that separate smart contract execution and consensus, and highlighting the differences with a table that shows how these different designs distinguish from one another. I can have better understanding after reading responses to review questions. However, the paper itself lacks the same level of clarity.

&

Thanks for pointing this out. We have added a table that shows the differences of these designs.

&

Table~\ref{tab:comparisons}

\\\hline

\textlabel{R3(5)}{row:r3_5}

&

Regarding experiment with injected faults. I understand that the current experiment evaluation mainly focuses on performance side. Other reviewer made suggestion that if some kind of experiments could be done to show that the system can satisfy safety related properties. I strongly agree with the suggestion. At least, some evaluation experiments may be feasible to demonstrate that the system could deliver these properties without a shadow of a doubt.

&

We did inject some random faults in our evaluation, 
they have no effect on either safety or performance.
This is because the fault-tolerance property: if faulty nodes is fewer than $f$, performance is unaffected.

&

\\\hline

\textlabel{R3(5)}{row:r3_5}

&

Inclusion of a subsection on limitations. Few currently unaddressed issues as the authors mentioned will be subjects of future work. A suggestion is to have a subsection that summarizes limitations  and restrictions of the current system, and future work.  The current design has a more restrictive smart contract model. Readers need to be aware of these restrictions and limits.

&

Thanks for this suggestion. We have added a subsection on limitations.

&

Section~\ref{sec:limitation}
\\\hline

\end{longtable}
\egroup
\twocolumn
\newpage
\else
\fi

\section{Introduction}

Blockchains make digital transactions possible without relying on a central authority.
One issue that hinders the wider deployment of blockchain-based applications such as cryptocurrencies~\cite{Bitcoin} and smart contracts~\cite{Ethereum}, is their low throughput and high latency.
This is partially due to the fact that {\em all} blockchain nodes are required to reach consensus on the order of transactions {\em and} execute them.

With the progress of the blockchain consensus algorithms~\cite{Byzcoin, Ouroboros, SnowWhite, Algorand, Thunderella, SBFT}, transaction execution will soon become a bottleneck. 
For example, {\em CryptoKitties}~\cite{CryptoKitties}, a popular game built on Ethereum blockchain, has clogged the network due to its complex genetic algorithm. This problem will be amplified by the computational demands of future smart contract applications.
A straightforward way to get rid of this bottleneck is dividing the blockchain nodes into groups (or shards) to process transactions in parallel~\cite{Dang, Luu2016, ZILLIQA, Omniledger, Chainspace, rapidchain, hyperledger}. 
However, existing approaches usually require extensive coordination but still suffer from congestion within the same group. 
\highlight{They also require a large group size, i.e., $3f'+1$ (or $2f'+1$~\cite{hyperledger}) nodes to tolerate $f'$ faults in each group.}
Additionally, they incur {\em livelocks} for smart contracts (cf. Section~\ref{sec:problem}).



%

\noindent{\bf Scalable execution.}
\highlight{In this paper, we propose $\paradigm$, a novel paradigm for scalable smart contract execution, by improving a traditional Byzantine fault-tolerance (BFT) architecture, called ``separating execution from consensus''~\cite{Yin2003}: 
\begin{enumerate}
    \item {\bf Parallel execution.} It distinguishes between {\em consensus nodes} and {\em execution nodes}. For simple transactions like cryptocurrency payments, consensus nodes 
    confirm them directly; for complex transactions that involve expensive execution, consensus nodes order them and assign them to different subsets of execution nodes ({\em {\Shard}s}) for parallel execution. 
    This can be seen as multiple instances of~\cite{Yin2003} running  in parallel.
    \item {\bf Asynchronous execution.} When complex transactions are executed by execution nodes, consensus nodes can keep processing simple transactions in a {\em non-blocking} way.
    This can effectively protect simple transactions from being starved by complex transactions.
\end{enumerate}
}
Compared with existing blockchain parallelization paradigms,
$\paradigm$ has the following advantages.
First of all, 
it supports asynchronous execution for smart contracts. 
Secondly, unlike prior sharding paradigm which requires extensive coordination among execution nodes,  
$\paradigm$ allows each individual execution node to execute the ordered transactions directly and independently.
Thirdly, it only requires $2f'+1$ nodes in each execution group, which can significantly reduce the group size\footnote{It only requires 70 nodes to reach a group failure probability of less than $10^{-6}$, whereas sharding schemes require 600 nodes.}.
Lastly, to the best of our knowledge, it is the {\em first} parallelization paradigm that is livelock-free for smart contracts.

We propose two ways to put $\paradigm$ into practice: on the existing Ethereum blockchain, and as a standalone blockchain.


\noindent{\bf $\paradigm$ for Ethereum.}
We apply $\paradigm$ to Ethereum~\cite{Ethereum} 
and show that we can make Ethereum support parallel and asynchronous contract execution {\em without introducing any hard-fork}. 
For a transaction that invokes a smart contract with expensive execution, 
the $\cons$ nodes (all Ethereum miners collectively serve as $\cons$ nodes) simply put it into the ledger 
without executing it, 
but they lock the states associated with this transaction and designate an {\Shard} for the execution.  
Once this transaction is confirmed, 
execution nodes in the designated group execute it off-chain and put the result into the ledger by making another transaction.
All these rules are enforced by the smart contracts themselves without changing the underlying consensus.

\noindent{\bf $\paradigm$ for a standalone blockchain.}
Following the $\paradigm$ paradigm, we propose a new public and permissionless blockchain called \name.
It leverages the state-of-the-art 
distributed randomness generation protocol~\cite{RandHund} to select a (rotating) committee of $\cons$ nodes which run a Byzantine consensus. 
The same randomness is used to construct groups of execution nodes.
Furthermore, 
$\name$ stores the whole blockchain into a distributed storage maintained by all nodes in the system, to support ``state sharding''.

Our contributions are summarized as follows:
\highlight{
\begin{itemize}
    \item We propose $\paradigm$, {\bf a paradigm for parallel and asynchronous} smart contract execution. 
    It supports a {\bf small group size of $2f'+1$} execution nodes and 
    it requires {\bf no coordination among execution nodes} and {\bf prevents (adversarial) livelocks}.
    \mbox{(Section~\ref{sec:paradigm})}
    \item We show how $\paradigm$ make Ethereum support parallel and asynchronous execution {\bf without introducing any hard-fork}. (Section~\ref{sec:ether})
    \item We propose \name, a new public, permissionless blockchain based on our proposed paradigm. It {\bf supports ``state sharding'' by further  separating storage from consensus}. (Section~\ref{sec:saberledger})
    \item We {\bf implement a prototype of $\name$} and deploy it on a network of \numprint{3467} nodes across 15 regions and 5 continents. 
    The results show that it can achieve a high throughput 
    even for complex transactions like Cryptokitties gene mixing, and it can effectively protect simple transactions from being starved by complex transactions. (Section~\ref{sec:eval})
    
\end{itemize}
}

\section{Background and Preliminaries}
\label{sec:background}

\subsection{Blockchains and smart contracts}
Blockchain technology has fueled a number of innovations such as cryptocurrencies~\cite{Bitcoin} and smart contracts~\cite{Ethereum}.
In particular, smart contracts permit execution of arbitrary code on top of blockchains.
However, blockchains introduce large overheads compared with traditional architectures.
For example, Bitcoin~\cite{Bitcoin} can only handle $\sim$7 transactions per second and each transaction requires one hour to be confirmed. 
Another reason is that every node in the system is required to execute all transactions.

Blockchains are usually permissionless, i.e., any node can join and leave at any time. 
Therefore, they need to be able to prevent sybil attacks.
PoW naturally provides {\em sybil-resistant identities}, since the number of sybils that an adversary can spawn is limited by its computing resources. 
Another solution for sybil-resistant identities is proof-of-stake (PoS), which limits adversary's power by its wealth.

Since blockchains are maintained in a distributed way, an upgrade to the blockchain-based software may lead to {\em hard-forks}: nodes running the old version software may see the transactions adhering to the new version as invalid.
During a recent hard-fork in Bitcoin, the network was divided into two separate parts~\cite{BitcoinCash}: Bitcoin and Bitcoin Cash~\cite{BitcoinCash}. 
Therefore, hard-forks have the risks of partitioning the committee.

\subsection{Separating execution from consensus}
\label{sec:separation}

Byzantine fault tolerant (BFT) state machine replication is a service where its state is replicated across $n$ servers and it can handle clients' requests as a single server. \atc{It ensures 
\begin{itemize}
\item {\em Safety}: all non-faulty servers execute the requests in the same order (i.e., consensus), and
\item {\em Liveness}: clients eventually 
complete their requests.
\end{itemize}
even if $f$ servers misbehave in arbitrary (``Byzantine'') ways.} 
One approach to build such services is {\em practical byzantine fault tolerance} (PBFT)~\cite{PBFT}, which requires $n=3f+1$ servers to tolerate $f$ faults.
In PBFT, one server, the {\em primary}, 
decides the order for clients' requests and forwards them to the other servers. 
Then, all servers agree on the order via a two-phase agreement to generate a {\em commit certificate} (CC), execute 
the requests and reply to the clients.
Clients wait for $f+1$ consistent replies to complete its request.

Yin et al. propose 
to split all servers in a BFT protocol into two clusters: an {\em agreement cluster} and an {\em execution cluster}~\cite{Yin2003}. 
The agreement cluster's job is to order clients' requests via a standalone BFT protocol (e.g., PBFT), 
send the ordered requests to the execution cluster, 
and relay replies to the clients.
In the execution cluster, $2f'+1$ servers are required to tolerate $f'$ faults, which is independent of the $f$ faults in the agreement cluster.


\subsection{Multisignatures and message aggregation}
\label{sec:multisignaure} 

A {\em multisignature} scheme allows multiple signers to produce a compact and joint signature on common input via an $\aggregate$ operation. Any verifier that holds the aggregated public key can verify the signature in constant time.
\atc{It can be easily built on top of BLS~\cite{BLS}, a well-known pairing-based signature scheme:
\begin{itemize}
    \item {\sf KeyGen}$(1^\lambda)$: output a key pair $(x, g^x)$, where $x$ is a random integer; and $g$ is a group generator;
    \item $\sign(M)$: output $\sigma := H(M)^x$, where $H()$ maps $M$ to a pseudorandom point on the elliptic curve;  
    \item $\aggregate(\sigma_1, ..., \sigma_n)$: output $\widetilde\sigma := \prod_{i=1}^n\sigma_i$;
    \item $\verify(\sigma, M)$: verify if $e(\sigma,g) = e(H(M), g^x)$; If the input is an aggregated signature, the public key should be aggregated as well: $g^{\widetilde x} := \prod_{i=1}^ng^x$.
\end{itemize}}
In practice, $\aggregate$ also outputs a bit map indicating which signers have (not) participated in the signing process, so that $\verify$ can compute the aggregated public key correspondingly. For the sake of brevity, we do not explicitly mention the bit map in the rest of the paper.

Multisignatures provide a useful property for {\em message aggregation}, which was used in ByzCoin~\cite{Byzcoin} to improve the scalability of PBFT.
\atc{The communication pattern of ByzCoin is shown in \fig~\ref{fig:ByzCoin} (in Appendix). }
Alternatively, hardware-assisted secret sharing~\cite{FastBFT} can achieve the same goal with smaller overhead but requires TEEs.

\subsection{Randomness beacon}
\label{sec:random_beacon}

Many recent blockchain consensus algorithms~\cite{Omniledger, Ouroboros, SnowWhite, Algorand, Dfinity} rely on a random beacon to generate randomness that is {\em unbiasable, unpredictable and third-party verifiable}. 
Such a random beacon is typically simulated by a distributed randomness generation protocol. 
Suppose there are $n$ nodes in the system and at most $f$ of them are malicious.
A commit-then-reveal~\cite{Ouroboros, Omniledger, RandHund} approach can be used to simulate a random beacon.  \atc{as follows: 
\begin{enumerate}
    \item Each node in the system
    \begin{enumerate}
        \item generates a random secret $r_i$, 
        \item publishes the commitment of $r_i$,
        \item splits $r_i$ into $n$ shares via threshold secret sharing,
        \item sends each share to each of other nodes together with a proof that the share is corresponding to $r_i$.
    \end{enumerate}
    \item All (honest) nodes reveal their shares.
    \item All (honest) nodes reconstruct all secrets, and the final random number is calculated as $r_1 \oplus ... \oplus r_n$.
\end{enumerate}}
An alternative approach is based on threshold signatures~\cite{Dfinity}, but it requires distributed key generation whenever the membership changes.

\section{Problem Statement}
\label{sec:problem}

\subsection{System setting and assumptions}
\label{sec:setting}

We target the setting of permissionless blockchains.
There are two types of entities in the system: {\em clients} and {\em nodes}. 
Clients 
issue {\em transactions} to transfer funds or run smart contracts. 
Nodes process clients' transactions via a blockchain consensus protocol. 
Notice that clients can play the role of 
nodes and vice versa.  
Each entity has a public/private key pair $(pk,sk)$ for digital signatures, and its identity is represented by its $pk$. 
\revisionmajor{\textlabel{
Following prior work}{r01:1}~\cite{BitcoinNG, Luu2016, Omniledger}, to against selfish mining attacks, we assume that at most 25\% nodes 
can fail {\em at any time point}.}
We also assume that messages 
can be delivered within a certain bound $\Delta$. 
All notations in this paper are listed in Table~\ref{notationtable}.

\begin{table}[ht]
\small
\centering
\scalebox{0.98}{
\begin{tabular}{|c|l|}
\hline
\textbf{Notation} & \textbf{Description} \\ \hline


$\Cli$       & client                      \\ \hline
$\Ser$       & server                      \\ \hline
$\order$      & $\cons$ node                     \\ \hline
$\exec$ & execution node  \\ \hline
$\trans$ & transaction \\ \hline
$\st$ & state \\ \hline
$B$ & transaction block \\ \hline
$R$ & result block \\ \hline



$n$ & number of $\cons$ nodes \\ \hline
$f$ &  number of faulty $\cons$ nodes \\ \hline
$m$ &  number of execution groups \\ \hline
$n'$ & number of execution nodes in each group\\ \hline
$f'$ &  number of faulty execution nodes \\ \hline
$v$  & view number  \\ \hline
$sn$ & sequence number \\ \hline


$H()$               &    cryptographic hash function \\ \hline

$pk/ sk$  & public / private key\\ \hline
$\sign() / \verify()$  & signature generation / verification\\ \hline
$\aggregate()$  & signature aggregation \\ \hline
$\sigma$     & signature \\ \hline
\end{tabular}
}
\caption{Summary of notations}
\label{notationtable}
\end{table}

\subsection{Parallel and asynchronous execution}
\label{sec:asyn}

Blockchain protocols are usually running in a sequential and blocking manner: a complex transaction (e.g., genetic algorithm in CryptoKitties) can congest the network so that simple transactions (e.g., cryptocurrency payments) cannot be confirmed on time.
\revision{\textlabel{Asynchronous execution}{r1:7} has been extensively used in web applications to improve performance and enhance responsiveness.
It enables some tasks to be executed separately from the main task and notify the main thread when the execution is completed~\cite{Asyn}}
In blockchain settings, {\em parallel and asynchronous execution} should satisfy the following requirements:
\begin{enumerate}
   \item complex transactions should be executed in parallel; 
    \item avoid blocking simple transactions by complex ones. 
\end{enumerate}

Blockchain researchers have already begun to investigate the possibility of integrating parallel execution with blockchains~\cite{Dang, Luu2016, Omniledger, Chainspace, rapidchain}. 
They divide the blockchain nodes into different groups to process transactions in parallel. 
However, these solutions require extensive coordination among blockchain nodes: 
they require BFT within each group, and two-phase lock/commit among different groups. 
Specifically, transactions that involve data objects in different groups must be committed in two-phases: lock the data first and access them afterwards.
If a transaction fails to acquire any of the locks, 
it releases all previously acquired locks and aborts. 
For each step of this two-phase protocol, every involved group needs to reach a Byzantine consensus.

\noindent{\bf Adversarial livelocks.} Even though the above approach can prevent deadlocks, it raises the rate of aborted transactions due to lock contention (called livelocks), because transactions will abort when they compete for the same lock.
Even worse, this problem also opens a channel for denial-of-service attacks: 
an adversary can easily abort other transactions by competing for locks. 
For example, two clients Alice and Bob share data objects $o_1$ and $o_2$ which are in two different groups $G_1$ (near Alice) and $G_2$ (near Bob) respectively.
Suppose Bob wants to make a transaction $\trans_B$ to update both $o_1$ and $o_2$ 
($\trans_B$ will first lock $o_2$ and then lock $o_1$). 
If Alice wants to make $\trans_B$ fail, she just needs to make a transaction $\trans_A$ to first lock $o_1$ and then lock $o_2$. 
In this case, both $\trans_A$ and $\trans_B$ will fail and Alice wins.
We name this attack {\em adversarial livelocks}.

	
		
	
	

\noindent{\bf Group size.} As each group runs a BFT protocol, they require $3f'+1$ nodes to tolerate $f'$ faults (in each group). 
Based on the analysis in~\cite{Omniledger}, each group requires at least 600 nodes to tolerate 25\% adversarial power: 
Suppose all execution groups are randomly chosen 
from an infinite pool of potential $\exec$s.
We use binomial distribution to calculate the probability that an execution group is {\em not} controlled by the adversary: 
\begin{equation}
P[f'<\lfloor\frac{n}{3}\rfloor] = \sum_{f'=0}^{\lfloor\frac{n}{3}\rfloor-1}\binom{n}{f'}\alpha^f(1-\alpha)^{n-f'}
\end{equation}
where $\alpha$=25\% is the adversarial power in the whole blockchain.
In order to get a system failure probability that is less than $10^{-6}$, it requires at least 600 $\exec$s in each group.
\revisionmajor{\textlabel{
The group failure probability is independent of the number of blocks being added to the chain.}{r02:1}
After a group being constructed, with a probability of $10^{-6}$, it is a ``faulty group'' (i.e., the number of failures is larger than $f$).
This probability will stay the same even with more blocks being added to the chain. }



\noindent{\bf Design goals.} To this end, we want to design a paradigm for {\bf parallel and asynchronous smart contract execution} with the following properties:

\begin{enumerate}
    \item {\bf minimized size for each execution group};
    \item {\bf no coordination among execution nodes};
    \item {\bf no (adversarial) livelocks}. 
\end{enumerate}

\begin{figure}[h]
    \centering
    \includegraphics[width=0.35\textwidth]{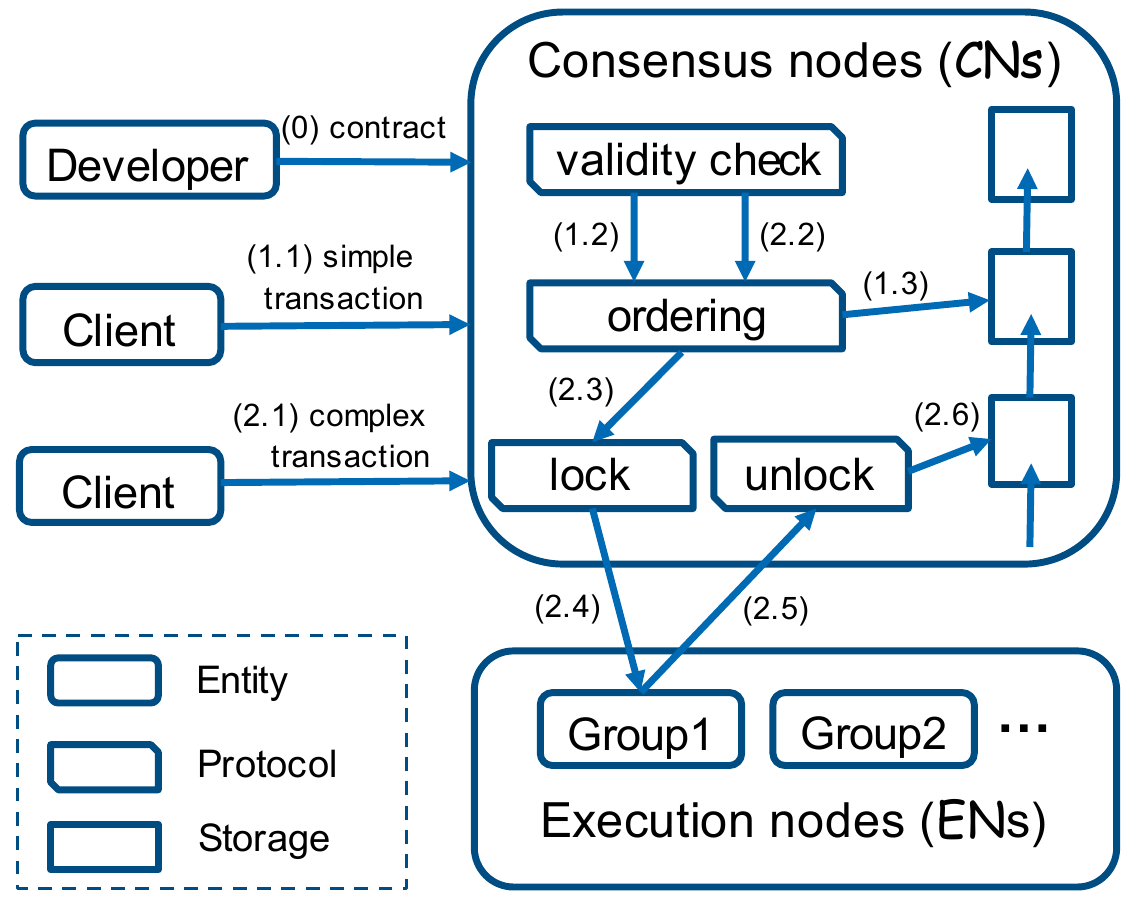} 
    \caption{Overview and workflow of $\paradigm$.}
    \label{fig:general}
\end{figure}

\section{\paradigm: Parallel and Asynchronous Smart Contract Execution}
\label{sec:paradigm}

In existing blockchains, 
transaction execution is tightly coupled with $\cons$.
We suggest that 
execution should be separated from $\cons$, 
which leads to $\paradigm$, a robust (e.g., livelock-free) and efficient paradigm for parallel and asynchronous smart contract execution.
\atc{(more benefits of separation will be discussed in Section~\ref{sec:discuss})}
\fig~\ref{fig:general} shows the basic architecture and workflow of $\paradigm$.
We distinguish between \textit{$\cons$ nodes} (denoted as $\order$s) and \textit{execution nodes} (denoted as $\exec$s);
and we also distinguish between simple transactions (e.g., cryptocurrency payments) and complex transactions (e.g., smart contract execution):
\begin{itemize}
    \item For a simple transaction, $\order$s
    \begin{description}
        \item[(1.1)] check its validity,
        \item[(1.2)] agree on its order, 
        \item[(1.3)] execute it (if needed) and update the blockchain;
    \end{description}
    \item For a complex transaction, $\order$s
    \begin{description}
        \item[(2.1)] check its validity,
        \item[(2.2)] agree on its order, 
        \item[(2.3)] lock its associated states,
        \item[(2.4)] assign it to an {\em execution group} and wait for the results, 
        (they can keep processing simple transactions while waiting)
        \item[(2.5)] collect the execution results,
        \item[(2.6)] unlock the states and update the blockchain.
    \end{description}
\end{itemize}
Notice that validity checking, transaction ordering and state locking can be done by $\order$s within one round of the underlying consensus protocol.

We leave it to the contract developers to decide whether a certain transaction should be simple or complex. 
A basic rule could be based on its execution time. 
Let $t_1$ be the latency of one consensus round,
$t_2$ be the execution time of this transaction,
$k$ be the number of transactions being batched in one consensus round (cf. Section~\ref{sec:saberledger}), 
and $m$ be the number of execution groups. If
\begin{center}
    $t_1 > \frac{k}{m}\cdot t_2$, 
\end{center}
$\exec$s will keep waiting for $\order$s. 
In this case, it is better to treat this transaction as a ``simple'' one.



Next, we explain how $\paradigm$ works. 
Recall that $\paradigm$ is a paradigm instead of a comprehensive protocol.
We simplify some details (e.g., we use transactions instead of blocks) for the ease of understanding. 
A comprehensive protocol for permissionless blockchains is discussed in Section~\ref{sec:saberledger}.

\subsection{Consensus nodes}
\label{sec:saber_con_and_exe}

The main job for $\order$s is to order transactions.
We assume that there are $m$ groups of $\exec$s (selected by an unbiased randomness, cf. Section~\ref{sec:saberledger}).
$\order$s maintain a separate and independent {\em sequence number} for each of them: $\langle sn_1, \dots, sn_m \rangle$.
After gathering $m$ {\em complex} transactions, $\order$s randomly
assign 
each transaction to an execution group and increase the corresponding sequence number $sn_i$.
Then, all $\order$s run the blockchain consensus to agree on $\langle \langle \trans_1, sn_1\rangle, \dots, \langle \trans_m, sn_m\rangle\rangle$;
and send each $\langle \trans_i, sn_i\rangle$ to the $i$th {\Shard}. 
After execution, each {\Shard} returns the execution result $\res_i$. 
In the end, $\order$s put $\langle \langle \trans_1, \res_1, sn_1\rangle, \dots, \langle \trans_m, \res_m, sn_m\rangle\rangle$ into the ledger. 
Notice that for {\em simple} transactions, $\order$ directly put them into the ledger after ordering.

Other than ordering, $\order$s have three additional jobs:

\noindent{\bf State maintenance.} 
The global state of the data ledger is maintained by $\order$s. 
Namely, $\order$s run 
a Byzantine consensus protocol to ensure the consistency and availability of the data ledger. 
Meanwhile, $\order$s leave the execution to $\exec$s and update the state based on the execution results.
$\exec$s can execute any transaction assigned to them, 
so that any transaction can be confirmed by one {\Shard} within one round, instead of being divided into multiple transactions~\cite{Omniledger, rapidchain}. 
This allows us to easily handle more complex transactions such as a smart contract calling other smart contracts.
On the other hand, this requires every $\exec$ has access to the data ledger as well. 
We solve this issue in Section~\ref{sec:saberledger}.
Notice that the transactions and results are only written to the data ledger {\em once}. 
The structure of the ledger is the same as other blockchain protocols like Bitcoin or Ethereum.

\noindent{\bf Lock handling.}
There may be multiple transactions aiming to access the same state. 
If these transactions are assigned to 
different {\Shard}s in one round, 
the state will diverge. 
We solve this problem by locking the state.  
Specifically, during the consensus round, 
$\order$s lock the states that this transaction wants to read/write. 
The locks are released only when the execution of this transaction is done.
Other transactions that want to access these locks need to wait for the next round. 
Since all the locks are handled by $\order$s, 
there are no livelocks in our paradigm.
A transaction locking all required objects gets executed and the locks will be released afterwards, i.e., no transaction can lock the acquired objects forever.
A fundamental difference between livelocks and our locking scheme is: 
livelocks cause all related transactions fail and the attacker pays no transaction fee;
in our case, an attacker locks all required objects can cause all other transactions fail, but she has to pay transaction fee.
\revision{\textlabel{Monetary penalty}{r2:6} is a common way to prevent denial-of-service attack.}
\revisionmajor{\textlabel{With monetary penalty}{r02:3}, a malicious client with resources to waste can still cause damage to the system.
We remark that denial-of-service attack can happen in any smart contract system if the attacker is willing to waste resources.
For example, in Ethereum, a malicious client with a fast network connection and willing to pay the gas fee, can always successfully call a smart contract and make other competing transactions fail.
However, its consistency property will always be maintained.
In this aspect, our scheme is equal to Ethereum.
}

The locks are specified by the contract developers (cf. Section~\ref{sec:ether}).
In particular, \fig~\ref{fig:shardcryptokitties} shows how consensus nodes find the objects touched by a transaction without executing it.
Developers are incentivized to reduce the gas consumption of their contracts.
However, it is clearly useful if we can provide assistants for lock handling at the compiler level, 
so that it is easier for the developers to develop their smart contracts.
We leave this as future work.

\noindent{\bf Validity checking.}
$\order$s are also responsible for checking the validity of the gathered transactions, e.g., whether a client has enough balance to make a payment, whether the required state of a transaction is locked etc.
A transaction will be ignored if it cannot pass the validity checking.
An alternative way of separation is to leave the validity checking to $\exec$s. However, this leads to a denial-of-service attack which is similar to livelocks.  
Suppose Bob wants to make a transaction $\trans_B$ to update an object $o_B$ and he is the only one who has write permissions.
Alice has a faster network connection and wants to delay the execution of $\trans_B$ for $k$ rounds.
Then, Alice just needs to issue $k$ transactions to update $o_B$ in front of $\trans_B$. 
\revision{\textlabel{In each round,}{r2:7}
only one of these $k$ transactions will be forwarded to $\exec$s and the rest will be cached.
In this case, $\trans_B$ has to wait for $k$ rounds until all Alice's transactions got rejected.
In contrast, if we have $\order$s check the validity, they will immediately find all these $k$ transactions are invalid and reject them in one round.
}


\subsection{Execution nodes}
\label{sec:exe}


It is enough to have $\exec$s 
execute $\trans_i$ directly if $sn_i$ is sequential to the sequence numbers they have seen.
This is based on the fact that, for each {\Shard}, $\order$s can never assign the same sequence number to different transactions due to the underlying consensus.
Therefore, we only need to make sure that the execution results returned by $\exec$s are correct. 
However, in each execution group, some $\exec$s may be faulty and they may return results that are different from the ones returned by correct $\exec$s. 
In this case, $\order$s need to resolve this dispute and decide which result to follow. 
In the rest of this section, we will introduce several existing solutions as well as our solution, and we will also provide a comparison.

\noindent{\bf Verifiable computation.}
{\em Verifiable computation} allows a {\em delegator} to outsource the execution of a complex function to some {\em workers}, and the delegator verifies the correctness of the returned result while performing less work than executing the function itself. 
The state-of-the-art solution for verifiable computation in cryptography is based on 
{\em succinct non-interactive argument of knowledge} (SNARK)~\cite{Pinocchio, SnarkforC}.
It allows the worker to provide a constant-size proof for the correct evaluation of a {\em circuit}. 
In these cases, each execution group only requires $(f'+1)$ $\exec$s (workers), because each $\exec$ can only crash but cannot return wrong results.
However, such solutions usually require a trusted setup, and the overhead for generating and verifying the proof is still too large to use in practice.


\noindent{\bf Trusted execution environments (TEEs).}
Another solution for verifiable computation is via 
TEEs~\cite{GP} (such as ARM TrustZone~\cite{AMD_trustzone} and Intel's SGX~\cite{SGX}), which provide protected memory and isolated execution so that 
adversaries can neither control nor observe the data being stored or processed inside them.
SGX also allows remote verifiers to ascertain the current configuration and behavior of a device via \emph{remote attestation}. 
Therefore, we can assume that each $\exec$ has a TEE and $\order$s only trust the results that are executed and signed by TEEs. 
Same as the zk-SNARK based solution, each execution group requires $(f'+1)$ $\exec$s.

\highlight{
\noindent{\bf Interactive verification.}
This solution was initially proposed by Canetti et al.~\cite{Canetti11}; later adopted by TrueBit~\cite{trubit} and Arbitrum~\cite{Arbitrum}. 
If two $\exec$s return different results, 
all $\order$s will collectively run as a judge and launch an
interactive verification game where they have one $\exec$ act as a solver and the other as a challenger.
The game proceeds in a series of rounds and each round narrows down the range of the execution in this dispute.
In each round, the challenger challenges a subset of the solver's execution,
and it challenges a subset of that set in the next round, 
until the judge can make a final decision on whether the challenge was justified.
In the end, either the cheating solver will be discovered and punished or the challenger will compensate for the resources consumed by the false alarm.
This solution introduces logarithmical number of rounds in terms the complexity of the function, and each round requires $\order$s to reach a consensus.
It requires at least one correct $\exec$ (and $f'+1$ in total) in each execution group to be the challenger.


}

\noindent{\bf Majority voting.}
We adopt the simplest way for dispute resolution.
Assuming honest majority ($2f'+1$) in each {\Shard}, if more than half of $\exec$s in each group return the same result, this result must be correct.  
Notice that $\order$s plus a single {\Shard} exactly match the architecture proposed by Yin et al.~\cite{Yin2003}:
 $\order$s correspond to the agreement cluster, 
and the group corresponds to the execution cluster. 
Even though we have multiple ``execution clusters'', the ``agreement cluster'' maintains a separate sequence number for each of them and there is no co-ordination among them. 
Therefore, this system can be considered as $m$ instances of the system proposed in~\cite{Yin2003} running in parallel. 
\revision{\textlabel{Following the analysis in}{r2:1} Section 2.2 (equation 3), it requires 70 $\exec$s in each execution group to reach a failure probability of $10^{-6}$.}



\section{Asynchronous Execution for Ethereum}
\label{sec:ether}

\atc{Ethereum recently announced that they will incorporate parallel execution via hard-forks~\cite{EtherShard}. 
In this section, we show how we can add parallel and asynchronous execution to Ethereum 
{\em without any hard-forks}. }


In this section, we show how we can add parallel and asynchronous execution to Ethereum 
{\em without any hard-forks}. 
We follow the architecture of $\paradigm$ (\fig~\ref{fig:general}): 
there are consensus nodes $\order$s and execution nodes $\exec$s.
We design a standard Ethereum smart contract for the $\paradigm$ execution management. 
Any Ethereum developer who wants to make their contract support our paradigm needs to include this contract and 
use the functionality exposed via its interface 
for their contract development. 
$\order$s are the original Ethereum miners collectively, 
and they are also allowed to register with the execution management contract and run as $\exec$s. 
That means the separation is only {\em in logic}. 
$\order$s run the standard Ethereum protocol as they are and the contract code will handle the execution.
Since we only require 70 $\exec$s to execute a transaction, the gas usage is much lower.
\revision{\textlabel{Notice that the inputs and outputs}{r2:2} of transactions are recorded in the blockchain state, and the code of smart contract is also publicly available. Therefore, new nodes that want to download and validate the entire chain can simply take the inputs, feed them into the contract code, and check if the execution results are consistent with those executed by $\exec$s. }

\remove{
Specifically, $\order$s:
\begin{enumerate}
    \item gather transactions,
    \item check their validity,
    \item lock the states,
    \item put them into a block,
    \item designate an execution committee for the execution of this block,
    \item put the block into the Ethereum blockchain via PoW.
\end{enumerate}
$\exec$s keep watching the blockchain. 
Once a block designated to them is confirmed (i.e., 12 confirmations), 
they execute the transactions in this block off-chain and update the global state by making another transaction.
}

\remove{
\begin{figure}[ht]
    \centering
    \includegraphics[width=0.45\textwidth]{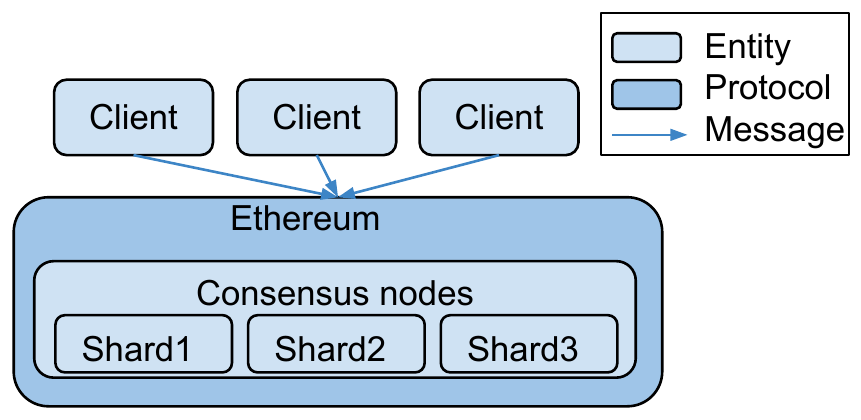} 
    \caption{\paradigm for Ethereum.\dawn{using a box to represent protocol is confusing}}
    \label{fig:ethereum}
\end{figure}
}

\subsection{Execution management}

\fig~\ref{fig:sharding_manager} shows the pseudocode of the execution management contract. 
We use a $pk$ list to record the identities of the nodes that have been registered with this contract as $\exec$s (line~\ref{3.1}).
Recall that each node is identified by its $pk$. 
and we assume that the $pk$s can be used to verify multisignatures (line~\ref{3.9}). 
\revision{\textlabel{The gas consumption}{r1:2} for signature verification is constant, independent of the complexity of the smart contract. So using signature verification instead of execution is worthwhile for complex smart contracts.}

Any Ethereum node can register as an $\member$ by calling the {\em Register} function (line~\ref{3.3}).
For sybil-resistance, we require $\exec$ to deposit some Ether to this contract account:
misbehaving $\exec$s will get punished in the same way as in proof-of-stake;
otherwise they will get some transaction fees as the miners.
$\exec$s are stored in the $pk$ list in an order according to the deposit they put, i.e., the one who deposits most will be in the head of the list.

$\exec$s are uniformly and periodically assigned to different {\Shard}s (line~\ref{3.2}) via the {\em Shuffle} function (line~\ref{3.5}).
After every epoch (e.g., \numprint{1000} confirmations), all $\exec$s (or a subset of them) jointly run a distributed randomness generation protocol 
off-chain to generate an unbiased random number $r$. 
Then, they input $r$ to the {\em Shuffle} function, 
which verifies $r$ first (recall that one property of this randomness is third-party verifiable),
re-assigns each $\exec$ to an execution group based on $pk$ and $r$ (line~\ref{3.7}).  
Note that $\exec$s who deposit more will also be assigned in the head of each execution group, i.e., {\em groups}[$i$][$0$] is the leader of {\em groups}[$i$].
Recall that each execution group requires ($2f'+1$) $\exec$s due to the requirement for majority voting.
\revision{\textlabel{In Bitcoin or Ethereum,}{r2:8} $\order$s not only make money from mining, but also from transactions fees. In our paradigm, we can distribute the transaction fees to $\exec$s.}

\begin{figure}[htbp]
\begin{mdframed}[leftline=false, rightline=false, linewidth=1pt, innerleftmargin=0cm, innerrightmargin=0cm]
\begin{algorithmic}[1]
\footnotesize
\State\textbf{contract} {\em ExecutionManager} 
\State\label{3.1}\phantom{----}$pk$[] $\members$
\State\label{3.2}\phantom{----}$pk$[][] $\shards$
\State\label{3.21}\phantom{----}$\trans$[][] {\em tasks}
\State\label{3.22}\phantom{----}{\em int} $sid$\Comment{initialized as 0}
\State\phantom{----}{\em int} $m$ \Comment{number of {\Shard}s}
 \\
\State\label{3.3}\phantom{----}\textbf{function} {\em Register}()
\State\label{3.4}\phantom{----}\phantom{----}add {\em caller}'s $pk$ to $\members$
\State\phantom{----}\textbf{end function}
\\
\State\label{3.5}\phantom{----}\textbf{function} {\em Shuffle}($r$)
\State\label{3.60}\phantom{----}\phantom{----}verify $r$; empty $\shards$
\State\phantom{----}\phantom{----}\textbf{for} each $\member$ in $\members$
\State\phantom{----}\phantom{----}\phantom{----}\label{3.7}$i\leftarrow H(r, \member)$ mod $m$; add $\member$ to $\shards[i]$

\State\phantom{----}\textbf{end function}
\\
\State\phantom{----}\textbf{function} {\em \revision{multisignature\_verify}}($i$, {\em M}, $\widetilde{\sigma}$)
\State\label{3.9}\phantom{----}\phantom{----} {\bf return} $\verify$($\shards[i], \widetilde{\sigma}, M$) 
\State\phantom{----}\textbf{end function}
\\
\State\phantom{----}\textbf{function} {\em \revision{signature\_verify}}($i$, $j$, {\em M}, ${\sigma}$)
\State\label{3.10}\phantom{----}\phantom{----} {\bf return} $\verify$($\shards[i][j], {\sigma}, M$) 
\State\phantom{----}\textbf{end function}
\State\textbf{end contract}
\end{algorithmic}
\end{mdframed}
\caption{Execution management for Ethereum.}
\label{fig:sharding_manager}
\vspace{-0.18em}
\end{figure}

\subsection{Running example: CryptoKitties}

We take CryptoKitties as a running example to explain how to use $\paradigm$ for parallel and asynchronous Ethereum contract execution without hard-forks.
CryptoKitties is a popular game built on the Ethereum blockchain~\cite{Kitties},  
which allows players to buy, collect, breed and sell digital cats.
\fig~\ref{fig:cryptokitties} shows the pseudocode of its contract with only a {\em giveBirth} function (adapted from~\cite{KittiesSource}), which runs an expensive gene mixing algorithm to create a new cat (line~\ref{4.8}). 
This complex genetic algorithm has clogged the Ethereum network recently: the number of unconfirmed transactions has remained consistently above \numprint{15000}~\cite{CryptoKitties}. 
Next, we show how to  improve the throughput by executing the {\em giveBirth} function in a parallel and asynchronous manner.

\begin{figure}[htbp]
\begin{mdframed}[leftline=false, rightline=false, linewidth=1pt, innerleftmargin=0cm, innerrightmargin=0cm]
\begin{algorithmic}[1]
\footnotesize
\State\textbf{contract} {\em CryptoKitties} 
\State\label{4.2}\phantom{----}{\em Kitty}[] {\em kitties}
\\
\State\label{4.3}\phantom{----}\textbf{function} {\em giveBirth}({\em matronID})
\State\label{4.4}\phantom{----}\phantom{----}{\em matron} $\leftarrow$ {\em kitties}[{\em matronID}]; check {\em matron}'s validity
\State\label{4.6}\phantom{----}\phantom{----}{\em sireID} $\leftarrow$ {\em matron.siringWithID}; {\em sire} $\leftarrow$ {\em kitties}[{\em sireID}]
\State\label{4.8}\phantom{----}\phantom{----}{\em childGenes} $\leftarrow$ {\em mixGenes(matron.genes, sire.genes)}\Comment{expensive operation}
\State\phantom{----}\phantom{----}{\em kitten} $\leftarrow$ {\em creatKitty(childGenes)}; add {\em kitten} to {\em kitties}
\State\phantom{----}\textbf{end function}
\State\textbf{end contract}
\end{algorithmic}
\end{mdframed}
\caption{Original CryptoKitties.}
\label{fig:cryptokitties}
\vspace{-0.18em}
\end{figure}

\fig~\ref{fig:shardcryptokitties} shows the $\paradigm$-version of CryptoKitties contract. 
It has a variable called $em$, which is initialized with the contract address of {\em ExecutionManager} (line~\ref{5.1}).
Therefore, we can directly use the {\em ExecutionManager} contract via $em$.
A transaction $\trans$ calling the {\em giveBirth} function will call the {\em giveBirth\_lock} function instead (line~\ref{5.4}).
$\order$s first check if the targeted matron has been locked (line~\ref{5.5}), i.e., being accessed by other transactions. 
If not, they check matron's validity (line~\ref{5.80}), e.g., whether it is a valid cat, whether it is pregnant, and whether its time has come. 
Then, they designate an idle {\Shard} for the execution of $\trans$ (line~\ref{5.80}). 
They also record the current block number, i.e., height of the current blockchain (line~\ref{5.11}). 
Next, they lock this matron by putting $\langle${\em matronID, groupID, blockNum,}$\trans\rangle$ into the {\em locks} array (line~\ref{5.12}), and put $\trans$ into the {\em tasks} array of the designated {\Shard} (line~\ref{5.13}).
Finally, they put this transaction into the ledger and update the state, as normal Ethereum miners.

Once $\trans$ is confirmed on the blockchain, 
each $\exec_i$ in the designated {\Shard} runs the {\em mixGenes} function off-chain, signs the result {\em childGenes}, and sends the signature $\sigma_i$ to the group leader {\em groups}[$i$][0].
The group leader combines the received $2f'+1$ signatures into a single multisignature $\widetilde{\sigma}$, and issues another transaction $\trans'$ calling the {\em giveBirth\_unlock} function. 
If other $\exec$s in {\em groups}[$i$] did not see $\trans'$ after a timeout, they send $\sigma_i$s to the second leader {\em groups}[$i$][1], so on and so forth, until $\trans'$ appears.

Upon receiving $\trans'$, $\order$s first check if $\trans$ has been confirmed  and verify the multisignature $\widetilde{\sigma}$ (line~\ref{5.17}).
Then, they add the new {\em kitten} to the {\em kitties} array (line~\ref{5.20}), unlock {\em matronID} (line~\ref{5.22}) and remove $\trans$ from {\em tasks} (line~\ref{5.23}).
Recall that all $\exec$s are acting as $\order$s as well. So all $\exec$s' states converge at this point even though they are in different {\Shard}s. 

\begin{figure}[ht]
\begin{mdframed}[leftline=false, rightline=false, linewidth=1pt, innerleftmargin=0cm, innerrightmargin=0cm]
\begin{algorithmic}[1]
\footnotesize
\State\textbf{contract} {\em SaberCryptoKitties} 
\State\label{5.1}\phantom{----}{\em ExecutionManager} $em$\phantom{----}\Comment{initialized with the contract address}
\State\label{5.2}\phantom{----}{\em Kitty}[] {\em kitties}
\State\label{5.3}\phantom{----}$\langle${\em int, int, int}$\rangle$[] {\em locks}\Comment{$\langle${\em kittyID, groupID, blockNum}$\rangle$}
\\

\State\label{5.4}\phantom{----}\textbf{function} {\em giveBirth\_lock}({\em matronID})
\State\label{5.5}\phantom{----}\phantom{----}{\bf if} {\em matronID} is in {\em locks}
\State\label{5.6}\phantom{----}\phantom{----}\phantom{----}cache $\trans$; {\bf return} \Comment{$\trans$ is the calling transaction}
\State\label{5.8}\phantom{----}\phantom{----}{\bf else}

\State\label{5.80}\phantom{----}\phantom{----}\phantom{----}check {\em kitties}[{\em matronID}]'s validity;  {\em groupID} $:=$ {\em em.sid}

\State\label{5.10}\phantom{----}\phantom{----}\phantom{----}{\em em.sid} $:=$ ({\em em.sid} + 1) mod $em.m$
\State\label{5.11}\phantom{----}\phantom{----}\phantom{----}{\em blockNum} $:=$ current\_block\_number
\State\label{5.12}\phantom{----}\phantom{----}\phantom{----}add $\langle$\revision{{\em matronID, groupID, blockNum}}$\rangle$ to {\em locks}
\State\label{5.13}\phantom{----}\phantom{----}\phantom{----}add $\trans$ to {\em em.tasks}[{\em groupID}]
\State\phantom{----}\textbf{end function}
\\
\\
\Comment{Nodes in {\em groups}[{\em groupID}] execute $\trans$ off-chain 
and generate a multisignature $\widetilde{\sigma}$ for $\langle${\em childGenes,}$\trans$$\rangle$}
\\
\State\label{5.14}\phantom{----}\textbf{function} {\em giveBirth\_unlock}({\em matronID}, {\em childGenes},  $\widetilde{\sigma}$)
\State\label{5.15}\phantom{----}\phantom{----}{\bf if} {\em matronID} is in {\em locks}
\State\label{5.16}\phantom{----}\phantom{----}\phantom{----}get {\em groupID}, {\em blockNum} and $\trans$ from {\em locks}
\State\label{5.17}\phantom{----}\phantom{----}\phantom{----}{\bf return} {\bf if} (current\_block\_number) - {\em blockNum} $<$ 13 {\bf or} $em$.{\em \revision{multisignature\_verify}}({\em groupID}, 
$\langle${\em childGenes},$\trans$$\rangle$,
$\widetilde{\sigma}$) 
is {\bf false}
\State\label{5.20}\phantom{----}\phantom{----}\phantom{----}{\em kitten} $\leftarrow$ {\em creatKitty(childGenes)}; add {\em kitten} to {\em kitties}
\State\label{5.22}\phantom{----}\phantom{----}\phantom{----}remove $\langle${\em matronID, groupID, blockNum}$\rangle$ from {\em locks}
\State\label{5.23}\phantom{----}\phantom{----}\phantom{----}remove $\trans$ from {\em em.tasks}[{\em groupID}]
\State\label{5.24}\phantom{----}{\bf end function}

\State\textbf{end contract}
\end{algorithmic}
\end{mdframed}
\caption{CryptoKitties in $\paradigm$ paradigm.}
\label{fig:shardcryptokitties}
\vspace{-0.18em}
\end{figure}


\section{\name}
\label{sec:saberledger}

\begin{figure}[h]
    \centering
    \includegraphics[width=0.45\textwidth]{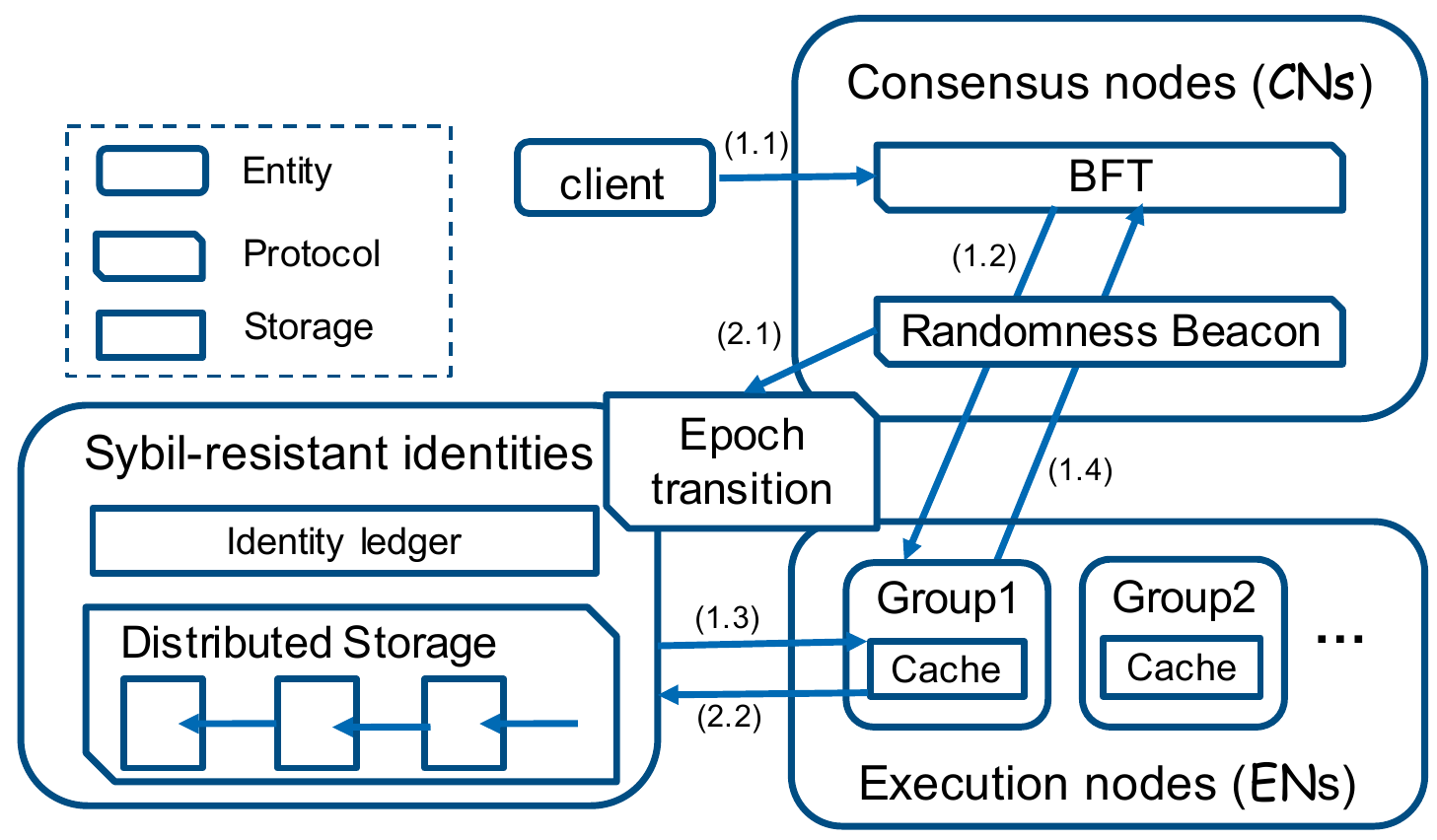} 
    \caption{Overview of $\name$.}
    \label{fig:saber}
\end{figure}

In this section, 
we propose a new public and permissionless blockchain called \name.
Its overview is shown in \fig~\ref{fig:saber}. 
\revision{\textlabel{Labels show the workflow:}{r3:10} the client first sends transactions to $\order$s (1.1);
$\order$s forward the transactions to $\exec$s (1.2);
$\exec$s read the state from the distributed storage (1.3) and return the execution results to $\order$s (1.4). After an epoch, $\order$s run randomness beacon the rotate $\order$s as well as $\exec$s (2.1); $\exec$s write back their cached sate to the storage (2.2).}
Overall, $\name$ follows the $\paradigm$ paradigm with the following augmentations:  
\begin{enumerate}
    \item batch processing by putting transactions into blocks;
    \item proof-of-stake (PoS) for sybil-resistant identities;
    \item BFT for the underlying consensus; 
    \item epoch transitions via a randomness beacon;
    \item ``state sharding'' via a distributed storage. 
\end{enumerate}

\subsection{Identity management and epoch transitions}

All nodes maintain a separate ledger called  {\em identity ledger} to record the sybil-resistant identities. 
One can get all the required $pk$s from the identity ledger.
This ledger can be implemented as a smart contract that is similar to \fig~\ref{fig:sharding_manager}. 
Any user can participate in $\name$ (i.e., become a sybil-resistant identity) by generating a key pair locally and making a deposit to this contract. 
Their identities are recorded in a $pk$ list.
The identity ledger also has a separated $pk$ lists to record $\order$s and groups of $\exec$s for all epochs.

We assume that an initial set of 
$\order$s was chosen in the bootstrapping phase of the system, 
and they run a distributed randomness generation protocol (i.e., randomness beacon, cf. Section~\ref{sec:background}) to generate an unbiased random number to build $m$ {\Shard}s, each of which has $n'=2f'+1$ $\exec$s. 
All participants are ranked according to the deposit they put, 
and the {\Shard}s are built in the same way as in Section~\ref{sec:ether} (line~\ref{3.7} in \fig~\ref{fig:sharding_manager}). 
Each {\Shard} has a leader which is the one who deposits most in that group.

To prevent an adaptive adversary from compromising more than a threshold number of $\order$s, as well as $\exec$s in each {\Shard}, we need to periodically rotate them from the underlying sybil-resistant identities.
After an epoch, $\order$s run the randomness generation protocol again to rotate $\order$s and $\exec$s.
The duration for each epoch depends on the required time for an adaptive adversary to compromise a node.
Following OmniLedger~\cite{Omniledger}, we only rotate a subset of nodes to minimize the chances of a temporary loss of liveness.
After rotating, they move to the next epoch.


\subsection{``State sharding'' via a distributed storage}
\label{sec:ipfs}

\atc{Recall that, in $\paradigm$, 
$\order$s are responsible for maintaining the global state of the blockchain.}
$\name$ further shards state by storing the blockchain in a distributed storage maintained by all nodes in the system. 
Specifically, all nodes run a distributed storage (e.g., IPFS~\cite{IPFS}) that only supports {\em read} and {\em write} operations. 
\atc{Notice that some distributed storage is not trusted, i.e., only provide availability and persistence, but not strong consistency. 
In this case,} $\order$s need to keep track of the version numbers and hashes of the last write operation for each state. 

Upon receiving a complex transaction associated with some states $\st$ (e.g., {\em matron} in \fig~\ref{fig:cryptokitties}), 
$\order$s only forward $\trans$ and the version numbers of $\st$ to $\exec$s, 
who read $\st$ from the distributed storage and make sure the version numbers match. 
Then $\exec$s execute the transaction and write the updated states $\st'$ back to the distributed storage,  
which accepts $\st'$ only if it has been signed by all $\exec$s in that {\Shard}.
$\exec$s will notify $\order$s when the writing is done. 
After receiving the notification from $\exec$s, $\order$s check the version numbers of the states from the storage and unlock the states if the version numbers are correct. 

\remove{
Upon receiving a complex transaction associated with some states $\st$ (e.g., {\em matron} in \fig~\ref{fig:cryptokitties}), 
$\order$s read $\st$ from the storage and send $\langle \trans, \st \rangle$ to $\exec$s.
\dawn{ENs should be able to read the relevant states from the distributed storage themselves, right? they don't need CNs to read them and send them to ENs.}
\Jian{Addressed. See the second paragraph below.}
After execution, $\exec$s return the updated states $\st'$ to $\order$s, who write them back to the storage. 
\dawn{need to update this part. CNs just need to store the hash of the updated states; and can store the states in the distributed storage.}\Jian{Addressed}
To reduce write frequency, $\order$s {\em cache} all the updated states and write them to the storage once. 
Namely, after every epoch, $\order$s run a consensus to agree on a set of updated states and write them back to the storage. 
}

To reduce write frequency, $\exec$s {\em cache} all the updated states and write them to the storage once. 
\revision{\textlabel{Naively, after execution,}{r1:4} the execution nodes need to write back the state to the storage immediately so that followup transactions access this state can be executed. However, this requires the execution nodes read and write the state very frequently, introducing a large overhead. To avoid this, we have the execution nodes cache the state, and have the consensus nodes forward the followup transactions to the execution group holding the corresponding state.}
In this case, $\order$s keep track of which {\st} have been assigned to which {\Shard}, and keep forwarding the transactions that are associated to that {\st} to the same {\Shard}. 
This works except that there is a transaction that is associated to {\st} and {\st'} which have been assigned to two different {\Shard}s. 
In this case, $\order$ will notify these two groups to write back {\st} and {\st'} to the storage.

\atc{
We can build a simple distributed storage 
by storing the states in the same way as prior sharding approaches~\cite{Omniledger, Chainspace, Eris}. 
Specifically, each shard keeps a subset of the global state. 
We assume there are $2f''+1$ nodes in each ``shard'' called {\em state nodes}. 
When $\exec$s need to read a state, they send a request to the corresponding shard.
All state nodes in that shard are required to response.
If a $\exec$ receives at least $f''+1$ consistent responses, it knows that the returned state is correct. 
When $\exec$s want to update a state, they 
generate a multisignature and send it to all state nodes in the corresponding shard.
Following the ``separating execution from consensus'' paradigm, the state nodes just need to update their states if the write requests are sequential. }

\subsection{Consensus and execution}

As in BFT, clients send transactions to the primary $\cons$ node $\order_p$ (cf. Section~\ref{sec:background}). 
If their transactions did not appear on the blockchain (in the distributed storage) after a timeout,
they send those transactions again to all $\order$s.

After gathering enough transactions, 
$\order_p$ puts the valid ones
into {\em transaction blocks}.
For simple transactions like cryptocurrency payments, 
$\order_p$ put them into a single transaction block $B$. 
Then, all $\order$s together run BFT 
to 
put $B$ into the distributed storage.

For each complex transaction $\trans_i$, $\order_p$ finds all its associated states $\st_i$ and 
locks them (in a way as shown in \fig~\ref{fig:shardcryptokitties}). 
Other transactions requiring $\st_i$ have to either be assigned to the same {\Shard} or wait for the next round.
Then, $\order_p$ puts all 
transactions arbitrarily and uniformly into $m$ transaction blocks $\langle B_1, \dots, B_m\rangle$ and assigns a sequence number $sn_i$ to each $B_i$. 
Recall that $\order$s maintain a separate and independent sequence number for each execution group.

$\order_p$ sends 
$\langle \langle B_1, sn_1\rangle \dots, \langle B_m, sn_m\rangle\rangle$ to all other $\order$s, 
who will check the validity of all $\langle\trans_i, \st_i \rangle$s and also check if there are multiple transactions in different blocks accessing the same state.
Then all $\order$s run BFT to agree on the proposal. 
In the end of the BFT round, they generate a commit certificate $CC$ (cf. Section~\ref{sec:background}) 
for each $\langle B_i, sn_i\rangle$. 

Next, $\order_p$ sends 
each $\langle B_i, sn_i, CC\rangle$ 
to all $\exec$s in the $i$th {\Shard}. 
To distribute the loads, 
the leader in each {\Shard} coordinates the communication between $\order_p$ and $\exec$s. 
Specifically, $\order_p$ only sends $\langle B_i, sn_i, CC_i\rangle$ to the group leader, who further distributes them to all $\exec$s in that group. 
Then, each $\exec$ reads the states from the distributed storage and executes the transactions in $B_i$ following the same order as they are being put into $B_i$, 
updates the corresponding states, 
and puts the updated states into a {\em result block} $R_i$.
It also generates a signature for $\langle R_i, sn_i \rangle$.
The group leader gathers signatures from $\exec$s 
and returns $\langle R_i, sn_i, \widetilde{\sigma}_i\rangle$ to all $\exec$s, where $\widetilde{\sigma}_i$ is a multisignature signaling that $\langle R_i, sn_i \rangle$ has been output by all $\exec$s in that {\Shard}.
$\exec$s write $\langle R_i, sn_i, \widetilde{\sigma}_i\rangle$ back to the distributed storage or their local cache,
and notify $\order_p$.
\atc{Recall that $\exec$s are ranked in each {\Shard} based on their deposits and group leader is the one who deposit most in that group.
If $\order_p$ receives no response after a timeout, it sends $\langle B_i, sn_i, CC\rangle$ again to the second leader in that group.
If $\exec$ notices a gap in the sequence numbers of the transaction blocks it has received, 
it first asks the group leader to fill the gap. 
If receiving no reply, it sends messages to all $\order$s.}




\revision{\textlabel{Recall that}{r1:3} we follow the same architecture as~\cite{Yin2003}: the agreement job of consensus nodes is to order clients’ requests via a standalone BFT protocol (e.g.,PBFT), send the ordered requests to the execution nodes, and relay replies to the clients. 
Then, $\order_p$ is exactly the BFT primary in~\cite{Yin2003}, and the view change procedure of BFT works for $\order$ as well. }

\remove{
Table~\ref{tab:comparisons} compares $\name$ with prior work.
The detailed comparison is discussed in Section~\ref{sec:related}.

\begin{table}[htbp]
\scriptsize
\centering
\scalebox{0.95}{
\begin{tabular}{|c|c|c|c|c|c|}
\hline
&\begin{tabular}[c]{@{}c@{}}Elastico\\\cite{ZILLIQA}\end{tabular} 
&\begin{tabular}[c]{@{}c@{}}OmniLedger\\\cite{Omniledger}\end{tabular} 
&\begin{tabular}[c]{@{}c@{}}Chainspace\\\cite{Chainspace}\end{tabular} 
&\begin{tabular}[c]{@{}c@{}}Eris\\\cite{Eris}\end{tabular} 
&$\name$ \\\hline
\begin{tabular}[c]{@{}c@{}}Support\\blockchains\end{tabular}            & $\surd$& $\surd$ & $\surd$   &X    &$\surd$\\\hline
\begin{tabular}[c]{@{}c@{}}Support\\cross-shard $\trans$s\end{tabular}  &  X     & $\surd$ & $\surd$   &$\surd$&$\surd$\\\hline
\begin{tabular}[c]{@{}c@{}}No \\ livelocks \end{tabular}                &  -     &  X      & X         &$\surd$&$\surd$\\\hline
\begin{tabular}[c]{@{}c@{}}No intra-shard \\ co-ordination\end{tabular}  & X      &  X      & X         &$\surd$&$\surd$\\\hline
\begin{tabular}[c]{@{}c@{}}Transparent \\ to clients\end{tabular}       & $\surd$&  X      & X   &X      &$\surd$\\\hline
\end{tabular}
}
\caption{Comparisons with other sharding protocols.}
\label{tab:comparisons}
\end{table}
}

\section{Implementation and Evaluation}
\label{sec:eval}

\subsection{Experimental setup} 

In order to systematically evaluate the performance of $\name$, 
we build a simulation framework allowing us to easily define the conditions and control the experiments. 

First, we setup a cluster of \numprint{3467} Amazon EC2~\texttt{t2.micro} VMs across 15 regions and 5 continents to introduce real network latency.
Each VM contains 1 2.3GHz vCPU, 1 GB memory and runs Amazon Linux 2.  

\highlight{
Second, we assign 70 $\exec$s to each execution group, which leads to a group failure 
probability of less than $10^{-6}$, base on the analysis in~\cite{Omniledger}. 
As a comparison,  existing sharding protocols require $\sim$600 nodes to reach the same failure probability, 
because they require at least $3f'+1$ nodes in each shard.
We only require $2f'+1$ nodes in each group.}
We can easily change the group size to make a trade-off between robustness and efficiency.

Third, we run SputnikVM~\cite{SputnikVM} -- a blockchain virtual machine for Ethereum -- on each $\exec$, 
so that we can test $\name$ with different types of Ethereum transactions ranging from simple cryptocurrency payments to CryptoKitties gene mixing algorithm.
Furthermore, we store the blockchain state in IPFS~\cite{IPFS}, but we cache the state as we discussed in Section~\ref{sec:ipfs}.

Last, recall that the bottleneck of $\name$ is its consensus layer. 
To this end, we make the consensus layer as a parameter as well.
We simulate the consensus layer by having it agree on a 1 MB block for every consensus round, with different rates: 
\begin{itemize}
    \item 0.1 rounds/s, which corresponds to the performance of current PoW-based or PoS-based consensus. 
    \item 1 rounds/s, which corresponds to the performance of current BFT consensus, e.g., PBFT~\cite{PBFT}, Byzcoin~\cite{Byzcoin}.
    \item 10 rounds/s, which conjectures the future of consensus protocols, e.g., EOS~\cite{EOS} and others. 
\end{itemize}
Assuming 1 MB block contains \numprint{2000} transactions (following Bitcoin and OmniLedger~\cite{Omniledger}), these consensus layers correspond to different throughputs of \numprint{200} TX/s, \numprint{2000} TX/s and \numprint{20000} TX/s\footnote{Notice that this throughput cannot be achieved by current public blockchains. For example, current CPU can only verify $\sim$\numprint{3500} ECDSA signatures per second. However, this figure is used to conjecture the future of consensus protocols.} respectively.
The advantage of this setup allows us to easily use different consensus layers.



To better compare $\name$ with previous state-of-the-art, we set three baselines for our benchmarks:
\begin{itemize}
    \item Throughput of current Ethereum, i.e., 30 TX/s at most.
    \item Throughput of {Ethereum-like} systems with different consensus layers, i.e., \numprint{200} TX/s, \numprint{2000} TX/s and \numprint{20000} TX/s respectively (for simple transactions).
    \item Throughput of current sharding protocols {with different consensus layers, i.e., \numprint{200} TX/s, \numprint{2000} TX/s and \numprint{20000} TX/s respectively (for simple transactions).}
\end{itemize}

\revision{\textlabel{Our evaluation}{r1:5} captures the setting with failures: if crash nodes is fewer than $f'$, performance is unaffected.
If it is more (or if a group cannot be formed), consensus nodes just stop forwarding requests to that group and wait for the next epoch to rotate groups. 
This only results in fewer groups.}
{\bf The worst case performance can be shown from the results of only one execution group.}
However, we assume this case rarely happens because faulty nodes will get punished.

\subsection{Evaluation results}



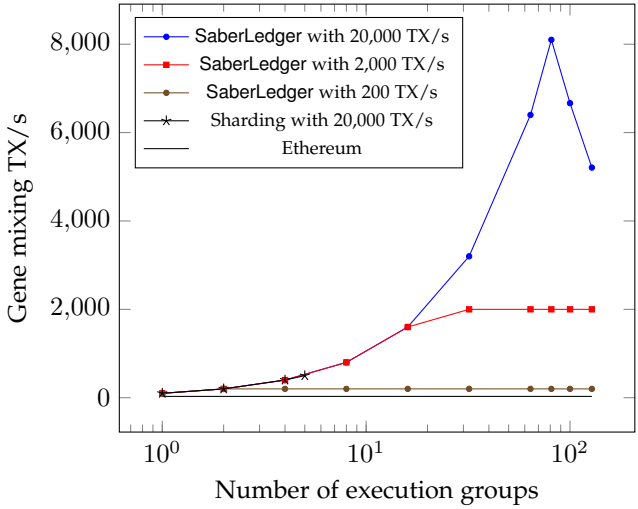
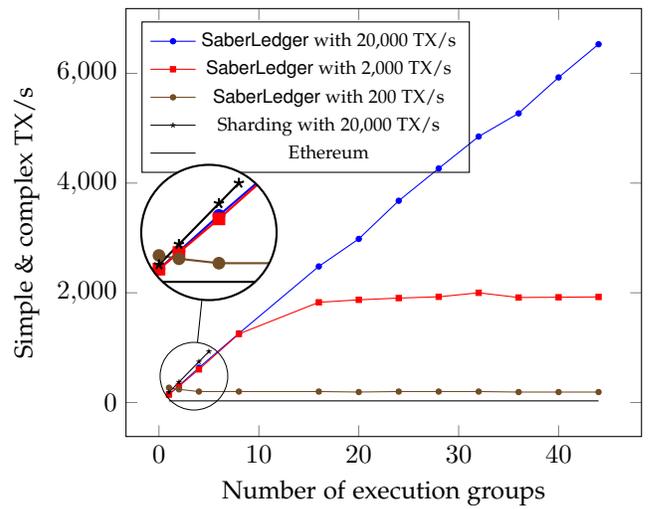
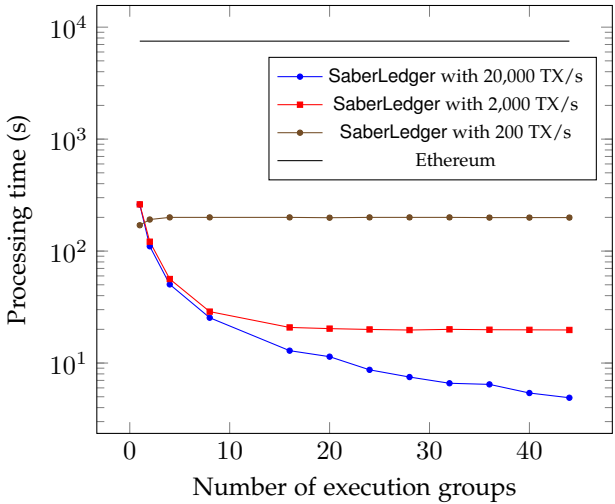
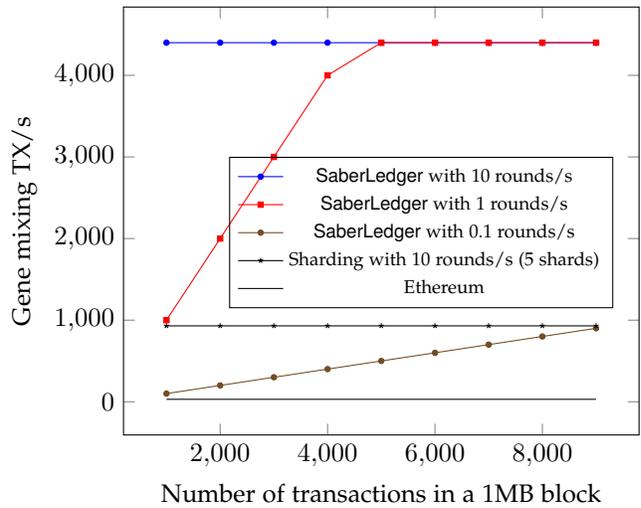
\begin{figure*}[ht!]
\centering
\subfigure[Throughput for gene mixing transactions vs. number of execution groups. (\numprint{2000} transactions in 1MB block).]{
\label{fig:saberledger_throughput}
        \begin{tikzpicture}[spy using outlines={circle, magnification=2, connect spies}]
\begin{axis}[
xmode=log,
xlabel={Number of execution groups},
ylabel={Gene mixing TX/s},
legend pos=north west,
legend style={font=\scriptsize}
]

\addplot+[mark options={scale=0.5}] coordinates {(1,100) (2,200) (4,400) (8,800) (16,1600) (32,3200) (64,6400)(81,8100)(100,6666)(128,5208)};
\addlegendentry{$\name$ with \numprint{20000} TX/s}

\addplot+[mark options={scale=0.5}] coordinates {(1,100) (2,200) (4,400) (8,800) (16,1600)  (32,2000) (64,2000)(81,2000)(100,2000)(128,2000)
 };

\addlegendentry{$\name$ with \numprint{2000} TX/s}

\addplot+[mark options={scale=0.5}] coordinates {(1,100) (2,200) (4,200) (8,200) (16,200) (32,200) (64,200) (81,200)(100,200)(128,200)};
\addlegendentry{$\name$ with \numprint{200} TX/s}


\addplot coordinates {(1,100) (2,200) (4,400) (5,500)}; 
\addlegendentry{Sharding with \numprint{20000} TX/s}



\addplot[mark=none] coordinates {(1,30) (2,30) (4,30) (8,30) (16,30) 
(32,30) (64,30) (81,30)(100,30)(128,30) };
\addlegendentry{Ethereum}

  
\end{axis}
   
\end{tikzpicture}
}
    ~
\subfigure[Throughput for real-world workloads with a mix of simple and complex transactions vs. number of execution groups. (\numprint{2000} transactions in 1MB block)]{
\label{fig:mix_throughput}
\begin{tikzpicture}[spy using outlines={circle, magnification=2, connect spies}]
\begin{axis}[
xlabel={Number of execution groups},
ylabel={Simple \& complex  TX/s},
legend pos=north west,
legend style={font=\scriptsize, style={fill=none}},
]

\addplot+[mark options={scale=0.5}] coordinates {(1,145.9293395) (2,307.9710145) (4,634.9206349) (8,1259.84252) (16,2480.620155) (20,2982.45614) (24,3678.16092) (28,4266.666667) (32,4848.484848) (36,5271.317829) (40,5925.925926) (44,6530.612245) };
\addlegendentry{$\name$ with \numprint{20000} TX/s}

\addplot+[mark options={scale=0.5}] coordinates {(1,145.4267126) (2,296.784831) (4,603.9076377) (8,1250) (16,1826.923077) (20,1871.921182) (24,1903.807615) (28,1926.977688) (32,2000) (36,1914.357683) (40,1919.191919) (44,1924.050633) };
\addlegendentry{$\name$ with \numprint{2000} TX/s}

\addplot+[mark options={scale=0.5}] coordinates {(1,270.3) (2,240.3) (4,200) (8,200) (16,200) (20,191.3) (24,200) (28,200) (32,200) (36,190.85) (40,190.8) (44,190.75) };
\addlegendentry{$\name$ with \numprint{200} TX/s}

\addplot+[mark options={scale=0.5}] coordinates {(1,186) (2,372) (4,744) (5,930)};
\addlegendentry{Sharding with \numprint{20000} TX/s}



\addplot[mark=none] coordinates {(1,30) (2,30) (4,30) (8,30) (16,30) 
(24,30) (28,30) (32,30) (36,30) (40, 30) (44,30)};
\addlegendentry{Ethereum}

\coordinate (spypoint) at (axis cs:3.5,480);
\coordinate (magnifyglass) at (axis cs:5,3100);
  
\end{axis}
\spy [black, size=1.8cm] on (spypoint)
   in node[fill=white] at (magnifyglass);
   
\end{tikzpicture}
}


\subfigure[Processing time for \numprint{50000} mixed transactions vs. number of execution groups. (\numprint{2000} transactions in 1MB block)]{
\label{fig:mix_time}
\begin{tikzpicture}
\begin{axis}[
ymode=log,
xlabel={Number of execution groups},
ylabel={Processing time (s)},
legend pos=north east,
legend style={font=\scriptsize, at={(0.97,0.88)}},
]

\addplot+[mark options={scale=0.5}] coordinates {(1,260.4) (2,110.4) (4,50.4) (8,25.4) (16,12.9) (20,11.4) (24,8.7) (28,7.5) (32,6.6) (36,6.45) (40,5.4) (44,4.9) };
\addlegendentry{$\name$ with \numprint{20000} TX/s}

\addplot+[mark options={scale=0.5}] coordinates {(1,261.3) (2,121.3) (4,56.3) (8,28.8) (16,20.8) (20,20.3) (24,19.96) (28,19.72) (32,20) (36,19.85) (40,19.8) (44,19.75) };
\addlegendentry{$\name$ with \numprint{2000} TX/s}

\addplot+[mark options={scale=0.5}] coordinates {(1,170.1812801) (2,191.4273824) (4,200) (8,200) (16,200) (20,198.6408782) (24,200) (28,200) (32,200) (36,199.1092481) (40,199.1614256) (44,199.2136304) };
\addlegendentry{$\name$ with \numprint{200} TX/s}

\addplot[mark=none] coordinates {(1,7500) (2,7500) (4,7500) (8,7500) (16,7500) 
(24,7500) (28,7500) (32,7500) (36,7500) (40,7500) (44,7500)};
\addlegendentry{Ethereum}

\end{axis}
\end{tikzpicture}
}
~
\subfigure[Throughput for gene mixing transactions vs. batch sizes (44 execution groups).]{
\label{fig:throughput_batchsize}
\begin{tikzpicture}
\begin{axis}[
xlabel={Number of transactions in a 1MB block},
ylabel={Gene mixing TX/s},
legend pos=north east,
legend style={font=\scriptsize, at={(0.95,0.65)}, style={fill=none}},
]

\addplot+[mark options={scale=0.5}] coordinates {(1000,4400) (2000,4400) (3000,4400) (4000,4400) (5000,4400) (6000,4400)
(7000,4400) (8000,4400) (9000,4400)};
\addlegendentry{$\name$ with 10 rounds/s}

\addplot+[mark options={scale=0.5}] coordinates {(1000,1000) (2000,2000) (3000,3000) (4000,4000) (5000,4400) (6000,4400) (7000,4400) (8000,4400) (9000,4400)};
\addlegendentry{$\name$ with 1 rounds/s}

\addplot+[mark options={scale=0.5}] coordinates {(1000,100) (2000,200) (3000,300) (4000,400) (5000,500) (6000,600)
(7000,700) (8000,800) (9000,900)};
\addlegendentry{$\name$ with 0.1 rounds/s}

\addplot+[mark options={scale=0.5}] 
 coordinates {(1000,930) (2000,930) (3000,930) (4000,930) (5000,930) (6000,930) (7000,930) (8000,930) (9000,930)};
\addlegendentry{Sharding with 10 rounds/s (5 shards)}

\addplot[mark=none] coordinates {(1000,30) (2000,30) (3000,30) (4000,30) (5000,30) (6000,30)
(7000,30) (8000,30) (9000,30)};
\addlegendentry{Ethereum}

\end{axis}
\end{tikzpicture}
}
    \caption{\label{fig:performance}Evaluation results.}
\end{figure*}

\noindent{\bf Performance with complex transactions.}
We first assume that all the workloads are complex transactions (i.e., Cryptokitties gene mixing), which gives us an estimate on the lower bound performance of $\name$.
We run the experiments with a varying number of execution groups and measure the peak throughput (TX/s) when the system is saturated. 
Note that each $\exec$ will receive a transaction block of size $\frac{1}{m}$ MB, where $m$ is the number of execution groups. 
Intuitively, the workload for each $\exec$ decreases as $m$ increases.
The results shown in \fig~\ref{fig:saberledger_throughput} validate this conjecture: 
as more execution groups being added, the performance of $\name$ keeps increasing until reaching the bottleneck of the consensus layer.
Specifically, for a fast consensus layer (\numprint{20000} TX/s, blue line), the throughput of $\name$ increases 
until reaching a throughput of \numprint{8100} TX/s, after which the signature verification becomes a bottleneck.
For a medium 
consensus layer (\numprint{2000} TX/s, red line), 
its throughput increases linearly until it reaches the bottleneck of its consensus layer when the number of the execution groups is 20.
For a slow consensus layer (\numprint{200} TX/s, brown line), its throughput is almost the same as its consensus layer. 
As a baseline, we also show the throughput of Ethereum, which is below 30 TX/s. 
In principle, the throughput of Ethereum should be similar with the brown line. 
However, in Ethereum, each block on average only batches \numprint{100} transactions due to the total gas limit for each block.
In $\name$, we can set a much higher gas limit and batch more transactions in one block, since each $\exec$ is only required to execute a subset of transactions. 
\noindent{\bf Remarks:} 
\begin{itemize}
    \item {\em $\name$ can achieve a high throughput even for complex transactions like Cryptokitties gene mixing.}
    \item {\em When there is no separation (the case of one execution group), even if the consensus layer is fast (\numprint{20000} TX/s), the throughput is still very low (\numprint{100} TX/s).}
    \item 
    {\em {Recall that sharding protocols require 600 nodes in each shard. 
    With \numprint{3467} nodes (5 shards), sharding protocols can reach a throughput of at most $\sim$\numprint{500} TX/s, even with a fast consensus layer (\numprint{20000} TX/s).
    With the same number of nodes and consensus layer, $\name$ can reach a throughput of \numprint{4201} TX/s. This demonstrates the prominent advantages of separating execution from consensus.}}
\end{itemize}

\noindent{\bf Performance with mixed workloads.}
In $\name$, simple transactions like cryptocurrency payments are confirmed asynchronously, independent of complex transactions. 
Therefore, the advantage of $\name$ will become more prominent if we consider real-world workloads that mix simple transactions with complex transactions.
To this end, we retrieve around \numprint{50000} transactions of recent \numprint{500} Ethereum blocks (from height \numprint{5998827} to \numprint{5999326}) from Etherscan~\cite{Etherscan}, 
and run these transactions on $\name$. 
To be conservative, we treat all contract invocations as complex transactions and assign them to different execution groups; 
and we treat cryptocurrency payments as simple transactions and confirm them directly in consensus layer.
We check if the sender or receiver address is a contract address by querying Etherscan’s API. 
Among these transactions, 47\% of them are simple and 53\% of them are complex. 
Furthermore, we treat two transactions as ``conflict'' as long as they are invoking the same contract, and one of them will be cached for the next round.
\fig~\ref{fig:mix_throughput} shows that the peak throughput of $\name$ for mixed transactions is significantly higher than only considering complex transactions (in \fig~\ref{fig:saberledger_throughput}).
For example, when the number of execution groups is 32, $\name$ can process another \numprint{1000} simple transactions in addition to \numprint{3200} complex transactions (for fast consensus).
\fig~\ref{fig:mix_time} shows that 
it takes 7s-7min for $\name$ to process all these \numprint{50000} transactions 
depending on the number of execution groups and consensus layer.
As a comparison, by inspecting the timestamps on the Ethereum blockchain, we found Ethereum requires 2 hours to finish processing these transactions.
\noindent{\bf Remarks:} 
\begin{itemize}
    \item {\em Asynchronous execution can effectively protect simple transactions from being starved by complex transactions, thus significantly improving the throughput.}
    \item {\em {In Ethereum-like systems or sharding protocols, complex transactions can block the processing of simple transactions. In the worst case, simple transactions have to wait until all complex transactions to be executed (at least 53 seconds). }}
\end{itemize}

\noindent{\bf Performance with a varying number of transactions in one 1MB block.}
As we mentioned, $\name$ can have a higher gas limit and batch more transactions in one block.
In principle, 1MB block can include around \numprint{9000} Ethereum transactions\footnote{For example, a Cryptokitties gene mixing transaction is \numprint{115} bytes.}. 
So the throughput of $\name$ can be improved if we batch more transactions in every block. 
To this end, we set both the number of execution groups and the group size as constants (44 execution groups) 
and run experiments with different batch sizes.
\fig~\ref{fig:throughput_batchsize} shows that, 
for a slow consensus layer (\numprint{200} TX/s, brown line), its throughput increases linearly as the batch size increases. 
For a medium consensus layer (\numprint{2000} TX/s, red line), its throughput increases linearly until it reaches the bottleneck of its execution layer (around \numprint{4400} TX/s).
For a fast consensus layer (\numprint{20000} TX/s, blue line), the throughput of $\name$ is exactly the same as its execution layer. 
{\bf Remarks:} 
\begin{itemize}
    \item {\em As the throughput for consensus layer increases, the execution layer becomes a bottleneck. 
    However, we conjecture that the blockchain network will become larger in the future. So we can introduce more execution groups.} 
    \item {\em For Ethereum-like systems and sharding protocols, increasing the batch size have no significant effect on throughput, as execution is the bottleneck and it blocks the processing.}
\end{itemize}

\section{Related Work}
\label{sec:related}

\noindent{\bf Hybrid consensus}
Another solution to avoid having all nodes execute all transactions is {\em hybrid consensus}~\cite{Byzcoin, pass2017hybrid, Solida}, which uses a slow permissionless blockchain protocol to bootstrap a fast permissioned blockchain protocol. 
For example in~\cite{Byzcoin}, a committee is elected by sliding a fixed-size window over a permissionless blockchain. 
Then, nodes in the committee run a BFT protocol 
to agree on the order of transactions {\em and} execute them, and other nodes just follow the results. 
They achieve Visa-level throughput for cryptocurrency payments. However, execution is still a bottleneck for smart contracts that require expensive executions. 


\noindent{\bf HyperLedger Fabric} 
Researchers in IBM propose an {\em execute-order-validate} paradigm for their permissioned blockchains~\cite{hyperledger}. 
In their paradigm, clients send transactions to multiple execution nodes  (called {\em endorsers}, which is specified by the smart contracts) first.
The endorsers execute the transactions independently and return the signed results (called {\em endorsement}) to the clients.
Each client collects endorsements until reaching the endorsement policy, 
and then submits them to a BFT-based ordering service, 
which establishes a total order on all endorsements and atomically broadcasts them. 
Compared with our consensus nodes, 
their ordering service is more generic: 
only does ordering but leaves the validation and ledger updates to the receivers.
\revision{\textlabel{This paradigm supports parallel execution,}{r3:1} but suffers from the same livelock issues as the sharding approaches: 
different endorsers may execute the same set of transactions in different order, 
in which case, all transaction fail as it requires multiple endorsers to produce the same result.}



\noindent{\bf ParBlockchain}
\revision{\textlabel{Amiri et al.}{r1:6}~\cite{ParBlockchain} propose a similar order-execute paradigm called OXII, based on which, they propose a permissioned blockchain called ParBlockchain.
However, they do not have multiple execution groups, instead, all execution nodes execute all transactions (cf. Figure 2 and 3 of~\cite{ParBlockchain}). 
As a result, SaberLedger has a much higher level of parallelism. 
Furthermore, in ParBlockchain, each execution nodes needs to multicast the execution results to all others, which introduce $O(n^2)$ communication complexity, whereas we have $O(n)$ communication complexity.
}

\noindent{\bf TrueBit and Arbitrum}
TrueBit~\cite{trubit} and Arbitrum~\cite{Arbitrum} also target the execution issues for smart contracts.
They also delegate the execution to a set of execution nodes and use interactive verification to resolve dispute (cf. Section~\ref{sec:exe}). 
As we discussed, the dispute resolution strategy in $\paradigm$ requires much less communication as well as coordination between consensus nodes and execution nodes.
In addition, $\paradigm$ further considers lock handling, 
which was ignored in TrueBit and Arbitrum.
Table~\ref{tab:comparisons} summarizes the comparisons between $\name$ and related work.

\begin{table}[htbp]
\scriptsize
\centering
\scalebox{0.95}{
\begin{tabular}{|c|c|c|c|c|c|}
\hline
&\begin{tabular}[c]{@{}c@{}}No \\ livelocks\end{tabular} 
&\begin{tabular}[c]{@{}c@{}}No intra-shard \\ co-ordination\end{tabular} 
&\begin{tabular}[c]{@{}c@{}}No inter-shard \\ co-ordination \end{tabular} 
&\begin{tabular}[c]{@{}c@{}}\# BFT \\ rounds\end{tabular} 
\\\hline

\begin{tabular}[c]{@{}c@{}}Sharding\\\cite{ZILLIQA}\cite{Omniledger}\cite{Chainspace}\end{tabular}  & X & X   &X    &3\\\hline

\begin{tabular}[c]{@{}c@{}}HyperLedger\\\cite{hyperledger}\end{tabular}      & X & $\surd$   &$\surd$ & 1\\\hline

\begin{tabular}[c]{@{}c@{}}ParBlockchain\\\cite{ParBlockchain} \end{tabular}                    &  -      & $\surd$         &-& 1\\\hline

\begin{tabular}[c]{@{}c@{}}TrueBit/Arbitrum\\\cite{trubit}\cite{Arbitrum}\end{tabular}        &  -      &  $\surd$          &-  & logarithm\\\hline

\begin{tabular}[c]{@{}c@{}}$\name$ \\ ~\end{tabular}        &  $\surd$      & $\surd$         &$\surd$&1\\\hline

\end{tabular}
}
\caption{\revisionmajor{Comparison with related work.}}
\label{tab:comparisons}
\end{table}

\section{Limitations}
\label{sec:limitation}

\revisionmajor{
Regardless of the various benefits brought by our paradigm, we have to admit that it has two limitations.
First, it changes the coding paradigm of smart contracts: the contract developers need to enumerate all dependencies when they develop the contract. It is clearly useful if we can provide
assistants for lock handling at the compiler level, so that it is easier for the developers to develop their smart contracts. We leave this as future work.
The second limitation is that the monetary counter-incentive can only alleviate denial-of-service attacks, instead of totally eliminating them.
}

\section{Conclusion}

In this paper, we propose a novel paradigm for parallel and asynchronous smart contract execution.
It neither requires extensive coordination nor suffers from (adversarial) livelocks, and it requires a small group size. 
We propose two ways to put this paradigm into practice.
We first apply it to Ethereum and show that we can make Ethereum support parallel and asynchronous execution without any hard-forks. 
Then, we propose a new public and permissionless blockchain $\name$, and show its performance by implementing a prototype.


\section*{Acknowledgments}

The work was supported in part by Zhejiang Key R\&D Plans (Grant No. 2021C01116, 2019C03133), National Natural Science Foundation of China (Grant No. 62002319, U20A20222) as well as a grant from China Zheshang Bank.

\bibliographystyle{plain}
\bibliography{main}


\begin{IEEEbiography}[{\includegraphics[width=1in,height=1.25in,clip,keepaspectratio]{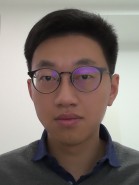}}]{Jian Liu}
is a ZJU100 Young Professor at Zhejiang University. Before that, he was a postdoctoral researcher at UC Berkeley. He got his PhD in July 2018 from Aalto University. His research is on Applied Cryptography, Distributed Systems, Blockchains and Machine Learning. He is interested in building real-world systems that are  provably secure, easy to use and inexpensive to deploy.
\end{IEEEbiography}

\begin{IEEEbiography}[{\includegraphics[width=1in,height=1.25in,clip,keepaspectratio]{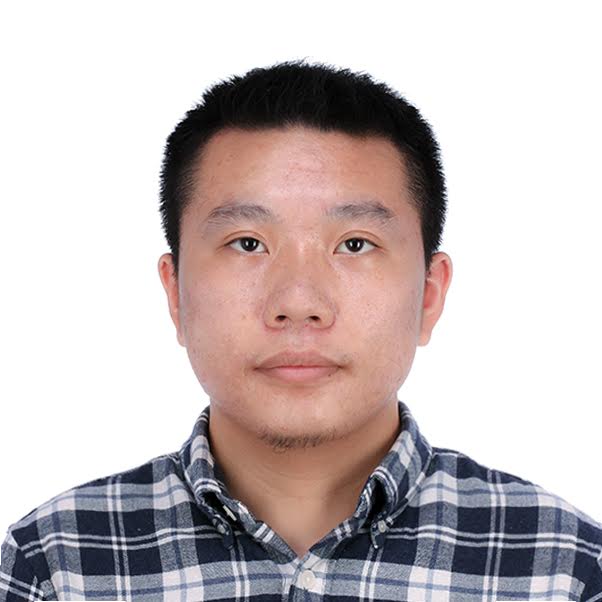}}]{Peilun Li}
received the BE degree from the Institute for Interdisciplinary Information Sciences, Tsinghua University, China, in 2015. He is currently working towards the PhD degree in the Institute for Interdisciplinary Information Sciences, Tsinghua University, China. His current research interests include distributed systems and blockchains.
\end{IEEEbiography}

\begin{IEEEbiography}[{\includegraphics[width=1in,height=1.25in,clip,keepaspectratio]{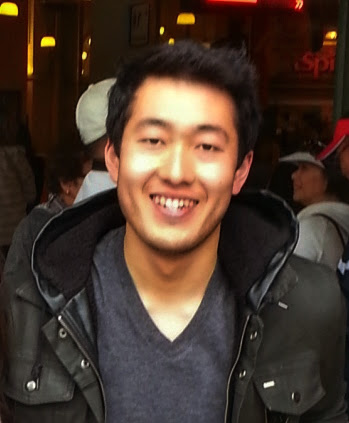}}]{Raymond Cheng}
is an Adjunt Faculty in the Department of Computer Science at the University of San Francisco.
Before that, he was a Co-Founder and CTO of Oasis Labs and a postdoctoral researcher at UC Berkeley. 
He got his PhD from University of Washington and M.Eng and B.S. from MIT.
He has contributed novel research papers in the areas of distributed systems and security,
most recently in the area of privacy-preserving systems that give users better control
over their data and spread digital freedom of speech and information to citizens around
the world.
\end{IEEEbiography}

\begin{IEEEbiography}[{\includegraphics[width=1in,height=1.25in,clip,keepaspectratio]{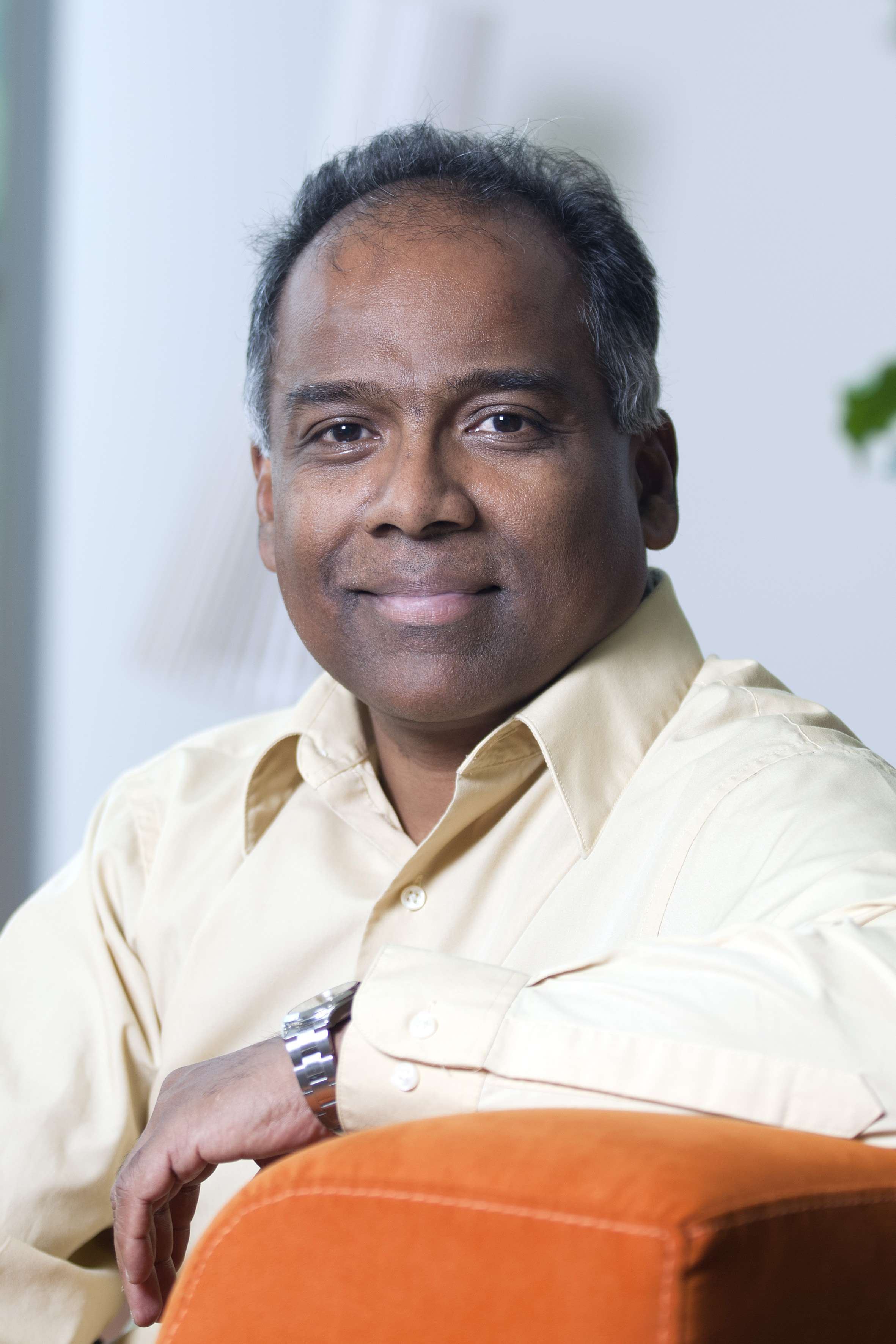}}]{N. Asokan}
is a Professor of Computer Science at the University of Waterloo (since 2019) where he holds a David R. Cheriton Chair. He is also an adjunct professor at Aalto University.
He was a Professor of Computer Science at Aalto University from 2013 to 2019 and at the University of Helsinki from 2012 to 2017. Between 1995 and 2012, he worked in industrial research laboratories designing and building secure systems, first at the IBM Zurich Research Laboratory as a Research Staff Member and then at Nokia Research Center, most recently as Distinguished Researcher.
He is an ACM Fellow and an IEEE Fellow.
\end{IEEEbiography}

\begin{IEEEbiography}[{\includegraphics[width=1in,height=1.25in,clip,keepaspectratio]{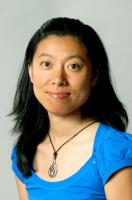}}]{Dawn Song}
is a Professor in the Department of Electrical Engineering and Computer Science at UC Berkeley. Her research interest lies in AI and deep learning, security and privacy. She is the recipient of various awards including the MacArthur Fellowship, the Guggenheim Fellowship, the NSF CAREER Award, the Alfred P. Sloan Research Fellowship, the MIT Technology Review TR-35 Award, and Best Paper Awards from top conferences in Computer Security and Deep Learning. She is an ACM Fellow and an IEEE Fellow. She is ranked the most cited scholar in computer security (AMiner Award). She obtained her Ph.D. degree from UC Berkeley. Prior to joining UC Berkeley as a faculty, she was a faculty at Carnegie Mellon University from 2002 to 2007. She is also a serial entrepreneur and has been named on the Female Founder 100 List by Inc. and Wired25 List of Innovators.
\end{IEEEbiography}

\end{document}